\documentclass[superscriptaddress,aps,twocolumn,prb,tightenlines,floatfix,showpacs,amsmath,amssymb,longbibliography]{revtex4-2}
\usepackage{graphicx}
\usepackage{dcolumn}
\usepackage{bm}
\usepackage[hypertexnames=false,colorlinks=true,linkcolor=blue,citecolor=blue,urlcolor=blue]{hyperref}
\usepackage{physics}
\usepackage{upgreek}

\begin{document}

\title{Electromagnetic Proximity Effects and Spontaneous Currents in Clean Superconducting Heterostructures}

\author{Jian-Lin Li}
\affiliation{Department of Electrophysics, National Yang Ming Chiao Tung University, Hsinchu, Taiwan}
\author{Chien-Te Wu}
\affiliation{Department of Electrophysics, National Yang Ming Chiao Tung University, Hsinchu, Taiwan}
\affiliation{Center for Emergent Functional Matter Science, National Yang Ming Chiao Tung University, Hsinchu, Taiwan}
\affiliation{Physics Division, National Center for Theoretical Sciences, Taipei, Taiwan}
\author{Klaus Halterman}
\affiliation{Independent Researcher, Cottonwood Heights, Utah, USA}

\date{\today}

\begin{abstract}
When ferromagnets are brought into contact with a superconductor, 
superconducting proximity effects give rise to a variety of interesting phenomena, 
including oscillatory singlet Cooper-pair amplitudes and long-ranged odd-frequency 
triplet correlations induced by the exchange interactions in the ferromagnets. 
From an electrodynamic perspective, however, it is equally important to understand 
when and how spontaneous currents can emerge. To investigate the interplay 
between electromagnetic and conventional superconducting proximity effects, 
we study clean ferromagnet/ferromagnet/superconductor spin-valve heterostructures 
in which the relative angle between the two ferromagnetic layers can be tuned. 
Our approach is based on a self-consistent numerical solution of 
the coupled Bogoliubov–de Gennes and Maxwell equations, providing 
a microscopic description capable of resolving physics on atomic length scales.
Several notable features emerge. In both the weak- and strong-exchange-field regimes, 
the central ferromagnetic layer plays a dominant role in generating 
sizable spontaneous currents when its exchange-field strength is varied. 
This suggests a practical route for maximizing electromagnetic proximity effects 
in future experiments. We further find that the resulting electromagnetic 
response extends across the entire superconducting layer, in sharp contrast to the ordinary inverse proximity effect, for which the penetration of magnetization into the superconducting layer is short-ranged.
In noncollinear configurations, 
the electromagnetic proximity effect also reconfigures the local magnetic-field orientation 
and reduces the angular mismatch between the fields in the two ferromagnetic layers. 
The long-ranged odd-frequency triplet amplitudes are consequently 
modified by the orbital response, which alters the underlying 
quasiparticle states by changing their momentum-space structure.
Finally, our framework can be naturally generalized
to other superconducting spintronic systems, possibly including Josephson junctions and 
altermagnet/superconductor heterostructures.
\end{abstract}

\maketitle

\section{Introduction}
\label{intro}
The superconducting proximity effect refers to the phenomenon in which a non-superconducting material exhibits superconducting properties when it forms a heterostructure with a superconductor. This effect was first observed in 1932 by Holm and Meissner~\cite{MeissnerHolm1932}, who connected two superconductors through a small metallic contact and observed zero-resistance behavior across the contact when an electric current was passed through it. Using the concept of Cooper pairs~\cite{Cooper_PhysRev.104.1189, BCS_PhysRev.106.162}, the proximity effect can be understood as the diffusion and penetration of the Cooper pair amplitude, $F_{\Delta}(\vb{r})=\langle\psi_{\uparrow}(\vb{r})\psi_{\downarrow}(\vb{r})\rangle$, from the superconducting material into the non-superconducting material. Throughout this article, the superconductors considered are assumed to be conventional {\it s}-wave superconductors. Therefore, in the following discussion, $F_{\Delta}(\vb{r})$ is also referred to as the singlet pairing amplitude. 
In a non-magnetic normal metal/superconductor (N/S) heterostructure, the Cooper pair amplitude in the N layer can be approximated by the asymptotic form $F_{\Delta}(\vb{r})\sim e^{-\abs{\vb{r}}/\xi_N}$ with $\abs{\vb{r}}$ measured from the N/S interface~\cite{parks1969superconductivity}. In the clean limit, the characteristic proximity length $\xi_N$ in the above asymptotic form is given by $\frac{\hbar v_F}{2\pi k_BT}$. Experimentally, the proximity length in the N layer has also been confirmed to be relatively long, on the order of micrometers~\cite{Mota1989}. Analyses of the proximity effect in the dirty limit can also be found in several seminal works~\cite{deGennes_RevModPhys.36.225, deGennes_physicsLetters_1963, parks1969superconductivity}.\par

The physical mechanism that allows the Cooper-pair amplitude to leak from the S side into the N side, thereby giving rise to the proximity effect, can be explained in terms of Andreev reflection~\cite{andreev1964reflection}. Through this mechanism, when an electron (or hole) incident from the N side has an energy within the superconducting gap, a hole (or electron) with the opposite spin is retroreflected at the N/S interface. That is, the Andreev reflection of an electron (or hole) is equivalent to the transfer of a Cooper pair into or out of the superconducting material.
This process results in a nonzero Cooper pair amplitude on the N side~\cite{Klapwijk2004, Pannetier2000}. Because $\xi_N$ can be relatively long, the proximity-induced superconducting condensate enables the Josephson effect~\cite{JOSEPHSON1962251} in an S/N/S junction, even with N-layer thicknesses reaching several hundred nanometers. Other important aspects of the Josephson effect in S/N/S structures are also discussed in Refs.~\cite{kulik1969macroscopic,kulik1972josephson,Likharev_RevModPhys.51.101,barone1982physics,Dubos_PhysRevB.63.064502}.\par

In contrast to the simple decay of the proximity effect in N/S heterostructures, ferromagnet/superconductor (F/S) heterostructures exhibit richer and more varied proximity behavior. Because the exchange interaction in ferromagnets tends to align spins of electrons in the same direction, while singlet Cooper pairs consist of electrons with opposite spins, the Cooper pair amplitude $F_{\Delta}(\vb{r})$ penetrating into the F layer is strongly suppressed compared with the N/S case.
In addition to its rapid decay, another notable feature of the Cooper-pair amplitude penetrating the F layer is its oscillatory behavior~\cite{Buzdin_RevModPhys.77.935, Demler_PhysRevB.55.15174, Halterman_PhysRevB.65.014509, Halterman_PhysRevB.66.224516}. This behavior arises because singlet Cooper pairs entering the F layer from the S layer acquire a nonzero center-of-mass momentum due to the exchange interaction. The superpositions of all these finite-momentum Cooper-pair states result in a sinusoidal oscillating amplitude~\cite{Demler_PhysRevB.55.15174}. 
The characteristic length $\xi_F$ describing the oscillation period can be qualitatively approximated as $\xi_F\approx(k_{F,\uparrow}-k_{F,\downarrow})^{-1}$, where $k_{F,\uparrow}$ and $k_{F,\downarrow}$ are the Fermi wavevectors of the spin-up and spin-down bands of the ferromagnetic material, respectively~\cite{Demler_PhysRevB.55.15174, Halterman_PhysRevB.65.014509, Halterman_PhysRevB.66.224516}. 
Assuming no Fermi-wavevector mismatch between the S and F layers, \(\xi_F\approx k_{F,S}^{-1}\left(\sqrt{1+I}-\sqrt{1-I}\right)^{-1}\), where \(I\equiv h/E_{F,S}\). Here, \(k_{F,S}\) and \(E_{F,S}\) are the Fermi wavevector and Fermi energy of the superconductor, respectively, and \(h\) is the strength of the exchange interaction in the ferromagnet.
In addition to describing the oscillations, $\xi_F$ also represents the characteristic decay length of $F_{\Delta}(\vb{r})$ in the ferromagnet, i.e., the proximity length~\cite{Halterman_PhysRevB.65.014509, Halterman_PhysRevB.66.224516}. Therefore, $\xi_F$ in a strong-ferromagnet/S system is expected to be much shorter than $\xi_N$ in an N/S system. In a strong-ferromagnet/S system, $\xi_F$ is typically on the order of a few nanometers and becomes even shorter in the half-metallic limit ($ I=1$).\par 

In F/S heterostructures, the aforementioned damped oscillating nature of the Cooper pair amplitude in the F layer is noteworthy and leads to various nontrivial consequences. First, in an F/S heterostructure, the superconducting transition temperature does not decrease monotonically with increasing F-layer thickness~\cite{Buzdin_RevModPhys.77.935}. This behavior arises from interference between the incident and reflected components of the oscillatory Cooper-pair amplitude within the F layer, bounded by the F/S interface and the outer surface. This results in thickness-dependent variations in Cooper-pair leakage from the S layer; these variations disappear once the F layer becomes sufficiently thick. 
This nonmonotonic variation of the transition temperature with increasing F-layer thickness may lead to a reentrant phenomenon: when the Cooper pair amplitude interference in the F layer is particularly strong, the transition temperature can vanish within a specific range of F-layer thickness and reappear at nonzero values with further increases in the F-layer thickness. This is both theoretically predicted~\cite{buzdin1990transition, Radovic_PhysRevB.44.759, Fominov_PhysRevB.66.014507} and experimentally observed~\cite{Garifullin_PhysRevB.66.020505, Zdravkov_PhysRevLett.97.057004, Zdravkov_PhysRevB.82.054517}. 

In addition to influencing the transition temperature, the oscillatory behavior of the Cooper pair amplitude allows multilayer structures composed of F/S interfaces to be classified into two main states, namely the so-called 0 phase and $\pi$ phase~\cite{Halterman_PhysRevB.69.014517,Barsic_PhysRevB.75.104502,Halterman_PhysRevB.70.104516,Ryazanov_PhysRevLett.86.2427, Kontos_PhysRevLett.89.137007, Oboznov_PhysRevLett.96.197003,Samokhvalov_PhysRevB.92.054511, Karabassov_PhysRevB.100.104502, Vargunin_ApplPhysLett.116.092601}. These two phases differ in the sign structure of the Cooper pair amplitude within the S layers: the 0 phase features identical signs across all S layers, whereas the $\pi$ phase is characterized by alternating signs between adjacent S layers. This characteristic is particularly important in constructing S/F/S-type Josephson junctions, where the Josephson current depends on the phase difference of the Cooper pair amplitude between the two superconductors. Thus, a $\pi$-phase junction yields a Josephson current with the opposite sign compared to a 0-phase junction. By adjusting the F-layer thickness, the phase of an S/F/S Josephson junction can be modified, allowing for improved design of desired SQUID elements. \par

Beyond suppressing and inducing spatial oscillations in the Cooper-pair amplitude, the exchange field in the F layer can also induce exotic spin-triplet correlations, even when the superconducting layer is a conventional {\it s}-wave superconductor. Odd-frequency pairing was first proposed by Berezinskii in the context of superfluid \(^3\)He~\cite{Berezinskii_JETPLett.20.287} and was later shown to arise in F/S heterostructures~\cite{Bergeret_PhysRevLett.86.4096,Volkov_PhysRevLett.90.117006,Halterman_PhysRevLett.99.127002, Halterman_PhysRevB.77.174511,RevModPhys.91.045005}.
Because the spin configurations of triplet state and 
$s$-wave spatial components are both
symmetric under particle exchange, the Pauli exclusion principle requires
the triplet correlations to be odd in the exchange of time. They therefore
vanish at equal times but may be finite for a nonzero time difference.
To express these correlations compactly, we define
$f_{\sigma\sigma'}(\vb{r},t,t^\prime)
\equiv
\langle
\mathcal{T}\psi_{\sigma}(\vb{r},t)
\psi_{\sigma'}(\vb{r},t^\prime)
\rangle$,
where $\mathcal{T}$ denotes the time-ordering operator. The triplet components are then given by
$f_0=(f_{\uparrow\downarrow}+f_{\downarrow\uparrow})/2$, $f_{+1}=f_{\uparrow\uparrow}$, and $f_{-1}=f_{\downarrow\downarrow}$.
These triplet correlations with total spin $S=1$ are clearly associated with the $m_S=0$ and $m_S=\pm1$ projections of the triplet pairing state.  For convenience, we define the linear combinations
$f_{\pm}=(f_{+1}\pm f_{-1})/2$
and use them throughout this paper. 
These odd-time components are commonly referred to as odd-frequency
spin-triplet correlations.
It is worth noting that not all F/S heterostructures support both types of odd-frequency spin-triplet correlations, $f_0$ and $f_{\pm}$. In a collinear magnetic configuration, the exchange field breaks full spin-rotation symmetry while preserving the spin projection along the magnetization axis, i.e., $[H,S_z]=0$. In this case, the $m_S=0$ component, $f_0$, is allowed, whereas the $m_S=\pm1$ components remain absent. 
If the magnetic configuration is noncollinear, no single global spin axis is conserved, and the equal-spin triplet components with $m_S=\pm1$ may also be induced~\cite{Halterman_PhysRevB.77.174511}.\par

While these triplet components emerge under different conditions, their most striking difference lies in how their odd-frequency correlations decay over space. Since the $f_0$ component consists of opposite-spin electron pairing, it behaves similarly to the conventional Cooper pair amplitude $F_{\Delta}(\vb{r})$ in ferromagnets, i.e., it remains short-ranged. The $f_{\pm}$ components, however, involve parallel-spin electron pairing, making them immune to ferromagnetic environments. As a result, they can penetrate ferromagnetic materials over longer distances than $F_{\Delta}(\vb{r})$ and $f_0$. As discussed earlier, long-range odd-frequency triplet correlations can emerge in F/S layered systems containing magnetic inhomogeneity. The topic has been extensively studied from both theoretical and experimental perspectives. In S/F/S Josephson junctions, a non-collinear magnetization in the ferromagnetic layers results in a notably slower decay of the critical current with increasing F-layer thickness~\cite{Robinson_Science.329.59}, relative to the collinear case. This can be regarded as indirect evidence for the long-range nature of the induced $f_{\pm}$ components, in contrast to $f_0$ and $F_{\Delta}(\vb{r})$. Another prominent example is the F1/F2/S trilayer structure, where the exchange fields in the F1 and F2 layers are misaligned by an angle $\theta$~\cite{Jara_PhysRevB.89.184502, Wu_PhysRevB.86.014523, Halterman_PhysRevB.92.174516, Wu_PhysRevB.90.054523, Wu_PhysRevB.98.054518, Montiel_PhysRevB.107.094513, Kamashev_PhysRevB.109.144517,PhysRevLett.109.237002}. 
When the layer thicknesses are fixed, the transition temperature depends
nonmonotonically on $\theta$ and reaches a minimum near the perpendicular
configuration~\cite{Jara_PhysRevB.89.184502, Wu_PhysRevB.86.014523}. This
behavior can be regarded as indirect evidence for enhanced long-range
$f_{\pm}$ correlations. In the systems considered, the configuration near
$\theta\simeq\pi/2$ favors the generation of equal-spin odd-frequency triplet
correlations, thereby enhancing the penetration of superconducting
correlations into the ferromagnetic layers and producing a stronger
suppression of the transition temperature. In addition to measurements of
$T_c$, the magnetic response has also been proposed as a probe of
odd-frequency triplet correlations. In particular, these correlations may
produce an anomalous paramagnetic Meissner response, providing a possible
signature of their odd-frequency character~\cite{Alidoust_PhysRevB.89.054508}.
In addition to the two prototypical structures that clearly exhibit long-range behavior, other configurations—such as F1/S/F2~\cite{Zhu_PhysRevLett.105.207002,Garifullin_PhysRevB.66.020505,Halterman_PhysRevB.72.060514,Halterman_PhysRevLett.99.127002,Halterman_PhysRevB.77.174511} and conical-F/S structures~\cite{Chiodi_EPL.101.37002, Wu_PhysRevB.86.184517, Wu_PhysRevLett.108.117005,Linder_PhysRevB.79.054523, Alidoust_PhysRevB.82.224504, Halasz_PhysRevB.84.024517}—also display various features associated with odd triplet superconductivity. Collectively, these ferromagnet/superconductor hybrid structures offer significant potential for spin-valve and spintronic applications.\par 

Thus far, the discussion has focused primarily on proximity effects in which
superconducting correlations penetrate from the superconducting component into
the nonsuperconducting region of a heterostructure. The adjacent ferromagnet,
however, can also produce a magnetic response inside the superconductor. One
such mechanism is the conventional inverse magnetic proximity effect, in which
spin-polarized superconducting correlations induce a local magnetization in the
superconductor near the F/S interface. This contribution is typically confined
to a region on the order of the superconducting coherence length $\xi_S$~\cite{Bergeret_PhysRevB.69.174504}.
Another form of reverse proximity effect, and the primary focus of the present work, is the electromagnetic proximity effect, which originates from the spontaneous equilibrium currents carried by proximity-induced superconducting correlations in the ferromagnetic layer.
Through Maxwell's equations,
this current distribution generates a magnetic field in the superconductor
that can extend over the electromagnetic screening length, which may greatly
exceed $\xi_S$~\cite{Mironov_ApplPhysLett.113.022601, Devizorova_PhysRevB.99.104519, Volkov_PhysRevB.99.144506}.

Experimentally, anomalous long-range magnetic responses and screening profiles
have been reported in several superconducting hybrid structures. In a
superconducting spin valve, a switchable magnetic moment was observed in a
normal-metal layer separated from the ferromagnetic elements by a thick
superconducting spacer~\cite{Flokstra_NatPhys.12.57}. Subsequent low-energy
muon-spin-rotation measurements on F/S thin films revealed an additional
screening contribution inside the superconducting layer that could not be
explained by the conventional Meissner response alone and was found to be
broadly consistent with the electromagnetic proximity mechanism~\cite{Flokstra_ApplPhysLett.115.072602, Stewart_PhysRevB.100.020505}.
Later measurements on Nb/Co-based structures identified contributions
consistent with both the spin-polarization and electromagnetic proximity
effects. However, the electromagnetic contribution was found to disappear upon
inserting only a few nanometers of a normal-metal spacer, and the observed
dependence on the magnetization direction showed notable deviations from the
existing theoretical predictions~\cite{Flokstra_PhysRevB.104.L060506}. 
Moreover, polarized-neutron reflectometry and detector-Josephson-junction
measurements on Ni/Nb bilayers found no detectable zero-field electromagnetic
proximity signal within the experimental resolution~\cite{Satchell_SupercondSciTechnol.36.054002}. 
These contrasting observations raise the central question of which
microscopic properties determine the existence, magnitude, sign, and spatial
profile of the electromagnetic proximity effect in a given F/S heterostructure.
 
Theoretically, the electromagnetic proximity effect was first formulated for
planar F/S structures in Ref.~\cite{Mironov_ApplPhysLett.113.022601}, where it
was shown that proximity-induced currents in the ferromagnet can generate a
long-range magnetic field inside the superconductor.
This framework was subsequently extended to superconducting spin valves,
where noncollinear magnetizations and long-range equal-spin triplet correlations
were shown to enhance strongly the induced electromagnetic response~\cite{
Devizorova_PhysRevB.99.104519}.
The interplay between spin-polarization and orbital contributions to the
magnetic response was further analyzed in
Ref.~\cite{Volkov_PhysRevB.99.144506}.
Further studies have considered
magnetic textures, superlattices, and the feedback of superconductivity on the
magnetic state~\cite{Bespalov_PhysicaC.595.1354032,Putilov_PhysRevB.105.064510,Kopasov_PhysRevB.110.214501,Kopasov_JSupercondNovMagn.38.241}. 
Most existing theoretical studies of the electromagnetic proximity effect
have employed quasiclassical Green-function approaches, such as the Usadel
or Eilenberger equations, coupled to Maxwell electrodynamics, sometimes
through an effective London-type current--vector-potential relation~\cite{
Mironov_ApplPhysLett.113.022601,Devizorova_PhysRevB.99.104519,Volkov_PhysRevB.99.144506,PhysRevLett.121.077002,Mironov_JETP_2021}. 
These methods have provided important insight into the long-range magnetic
response of F/S heterostructures.
However, by construction, quasiclassical approaches integrate out the full
microscopic quasiparticle structure and typically represent interfaces through
matching conditions rather than resolving the quasiparticle wave functions
across them.

A central difficulty in treating the electromagnetic proximity effect is that
the equilibrium current depends on the vector potential, while the vector
potential must itself be determined from the current and magnetization through
Maxwell's equations. The superconducting order parameter, 
quasiparticle amplitudes, current 
density, induced magnetization, and electromagnetic field are therefore
mutually coupled.
Existing treatments have often reduced the complexity of this coupled problem
by using effective current--vector-potential relations, scale-separation
approximations, or London-type descriptions of the long-distance
electromagnetic response. As a result, the superconducting and electromagnetic
degrees of freedom have generally not been determined simultaneously within a
fully microscopic self-consistency loop.

\begin{figure}[ht]
    \centering
    \includegraphics[width=.85\linewidth,clip]{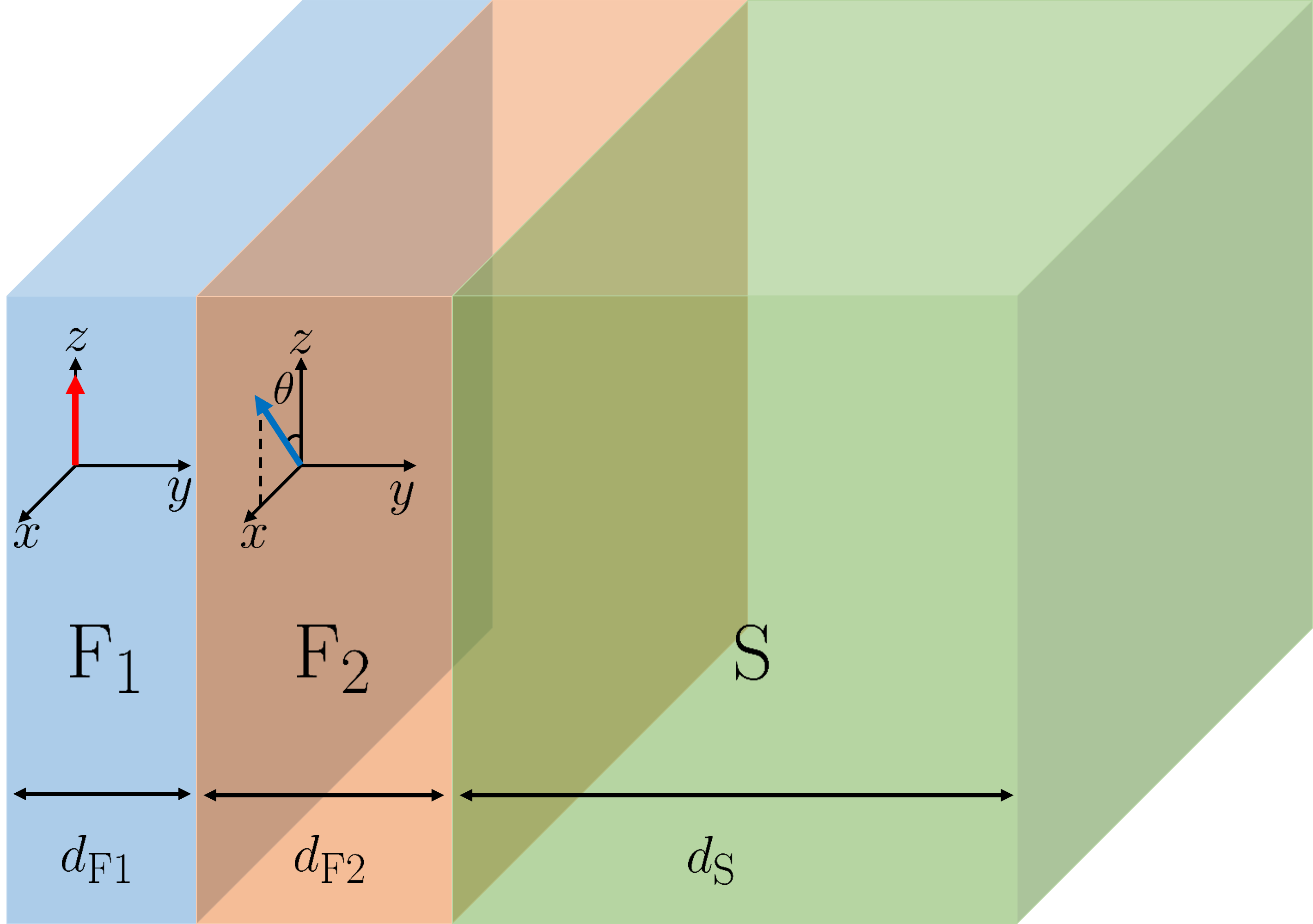}
    \caption{Schematic illustration of the F1/F2/S heterostructure composed of ferromagnetic and superconducting layers. The $y$-axis is normal to the interfaces. The first (second) ferromagnetic layer, denoted F1\,(F2), has an in-plane homogeneous magnetization or exchange field oriented at an angle of $0\,(\theta)$ relative to the $z$-axis in the $x$–$z$ plane.}
    \label{fig_model}
\end{figure} 

In the present work, we instead solve the continuum Bogoliubov--de Gennes (BdG) and
Maxwell equations self-consistently. Our approach is closely related in spirit
to the self-consistent BdG--Maxwell treatment developed for superconducting
vortices in Ref.~\cite{Gygi_PhysRevB.43.7609}. By exploiting the symmetries and
boundary conditions of the planar F/S geometry, we determine the
superconducting order parameter, quasiparticle amplitudes, equilibrium current, 
magnetization, vector potential, and magnetic-field profile on the same
microscopic footing. This framework allows us to examine how microscopic parameters and geometries
control the electromagnetic proximity effect and to clarify the origin of the
different responses observed across F/S heterostructures.
The remainder of this paper is organized as follows. In Sec.~II, we introduce
the theoretical framework and numerical methods. In Sec.~III, we present and
discuss the results. Finally, we summarize our main findings and conclude in
Sec.~IV.\par

\section{Model and method}
\label{methods}
In this section, we begin by considering a standard spin-valve type F1/F2/S heterostructure, schematically illustrated in Fig.~\ref{fig_model}, and focusing on how to self-consistently treat the electromagnetic proximity effect. The S region is assumed to be a conventional 
s-wave superconductor and the relative angle between the in-plane magnetizations of F1 and F2 is denoted by $\theta$. The system is assumed to be homogeneous along the $x$ and $z$ directions and
finite along the $y$ direction. Specifically, the dimensions in the $x$--$z$
plane are taken to be sufficiently large that boundary effects and spatial
variations along these directions can be neglected, while the S, F1, and F2
layers have thicknesses $d_S$, $d_{F1}$, and $d_{F2}$, respectively.

For completeness, the formulation also allows for a spatially uniform external magnetic field directed in the $x$–$z$ plane.
Accordingly, all physical observables are expected to depend only on the $y$ coordinate, with no variation along $x$ and $z$. Within the framework of the grand-canonical ensemble, the system can be described by the following effective Hamiltonian, expressed in terms of the electron field operators, using  Gaussian units and the charge of an electron $e=-\abs{e}<0$:
\begin{widetext}
    \begin{align}
        H_{\text{eff}}
        &=
        \int d^3r\sum\limits_{s=\uparrow,\downarrow}\psi^\dagger_s(\vb{r})
        \left[
            \frac{\left(-i\hbar\grad-\frac{e}{c}\vb{A}(\vb{r})\right)^2}{2m^*}
            +U(y)-E_F
        \right]
        \psi_s(\vb{r})
        +
        \frac{1}{2}\int d^3r\sum_{s,s^\prime}
        \left[
            \Delta(\vb{r})\left(i\sigma_y\right)_{ss^\prime}\psi^\dagger_s(\vb{r})\psi^\dagger_{s^\prime}(\vb{r})+h.c.
        \right]\nonumber
        \\
        &\quad-
        \int d^3r\sum_{s,s^\prime}\psi^\dagger_s(\vb{r})\big[\vb{h}(y)\vdot(-\vb*{\sigma})\big]_{s,s^\prime}\psi_{s^\prime}(\vb{r})
        -
        \int d^3r\,\left(-\frac{g^*\mu_B}{2}\right)\sum\limits_{ss^\prime}\psi^\dagger_{s}(\vb{r})\big[\big(\vb{B}_{\text{ex}}+\vb{B}_{\text{mat}}(y)\big)\vdot\vb*{\sigma}\big]_{ss^\prime}\psi_{s^\prime}(\vb{r}), \label{Hamiltonian}
    \end{align}
\end{widetext}
where $m^*$ and $g^*$ denote the effective mass and {\it g}-factor of the electron (not the complex conjugations of $m$ and $g$).\par 

For simplicity, we assume that there is no Fermi wavevector mismatch between different layers. The term $U(y)$ represents a spin-independent scattering potential arising at the interfaces of adjacent layers. To focus on the electromagnetic proximity effect, we choose to neglect the effect of $U(y)$ by setting $U(y)=0$ in this work. In Eq.~\eqref{Hamiltonian}, $\Delta(\vb{r})$ and $\vb{A}(\vb{r})$, treated self-consistently within our framework, denote the superconducting pairing potential and the total vector potential of the system, respectively. $\vb{B}_{\text{ex}}$ and $\vb{B}_{\text{mat}}$ denote the magnetic fields
arising from the external configuration and from magnetic sources within the
material, respectively. The latter include the electron spin magnetic moments
and electric currents.
Based on the symmetry discussed above, $\vb{B}_{\text{mat}}$ is expected to
depend only on $y$ and to lie entirely within the $x$--$z$ plane.
Accordingly, Eq.~\eqref{Hamiltonian} is written in terms of $\vb{B}_{\text{mat}}(y)$ rather than $\vb{B}_{\text{mat}}(\vb{r})$.
In Eq.~\eqref{Hamiltonian}, all electromagnetic quantities are treated as c-numbers rather than operators; thus, the model corresponds to a semiclassical treatment of the electromagnetic field. The matrices $\sigma_0$ and $\vb*{\sigma}$ denote the identity matrix and the Pauli matrices in spin space. The term $\vb{B}_{\text{mat}}(y)\vdot\vb*{\sigma}$ can be regarded as a mean-field treatment of spin–spin and spin-current interactions mediated through the magnetic field $\vb{B}_{\text{mat}}$.
Finally, the vector function $\vb{h}(y)$ represents the exchange fields in the F layers, which directly influence the intrinsic local magnetization. 
Note that in our convention, $\vb{h}(y)$ points essentially in the same direction as intrinsic magnetizations in the F layers.
It lies entirely in the $x$-$z$ plane and its expression is given by $\vb{h}(y)=(h_x(y),0,h_z(y))$, where $h_x(y)=h_2\sin\theta\vb{1}_{\text{F2}}(y)$ and $h_z(y)=h_1\vb{1}_{\text{F1}}(y)+h_2\cos\theta\vb{1}_{\text{F2}}(y)$.
Here the indicator function of region $A$, $\vb{1}_{A}(y)$, is defined by
\begin{equation}
\vb{1}_{A}(y)=
\begin{cases}
1, & y\in A,\\
0, & y\notin A.
\end{cases}
\end{equation}\par

In the following, we formulate the problem within the BdG framework, coupled self-consistently to Maxwell’s equations and constrained by the relevant physical conditions and symmetries.
In this problem, $\Delta(\vb{r})$ and $\vb{A}(\vb{r})$ are the central
quantities to be determined self-consistently.
Although $\vb{B}_{\text{mat}}(y)$ also requires iterative updates, it is relatively straightforward to handle.
The superconducting order parameter is defined as $\Delta(\vb{r})\equiv-\mathcal{V}_{\text{eff}}(y)\langle\psi_\downarrow(\vb{r})\psi_\uparrow(\vb{r})\rangle=\mathcal{V}_{\text{eff}}(y)\langle\psi_\uparrow(\vb{r})\psi_\downarrow(\vb{r})\rangle$ where $\mathcal{V}_{\text{eff}}(y)=\mathcal{V}_{\text{eff}}\vb{1}_{\text{S}}(y)$ with $\mathcal{V}_{\text{eff}}>0$ denoting the BCS singlet coupling strength in the S layer. The explicit form of the total vector potential $\vb{A}(\vb{r})$ will be discussed later in conjunction with Maxwell’s equations. Our problem can be simplified at this stage by exploiting the gauge symmetry (or gauge transformation) of the system. The gauge transformation leads to the following relations:
\begin{subequations}
    \begin{align}
        \psi_{s}(\vb{r})&\mapsto\psi_{s}^{\text{new}}(\vb{r})=\psi_{s}(\vb{r})e^{\frac{-i\abs{e}}{\hbar c}\chi(\vb{r})},\\
        \vb{A}(\vb{r})&\mapsto\vb{A}^{\text{new}}(\vb{r})=\vb{A}(\vb{r})+\grad\chi(\vb{r}),\\
        \Delta(\vb{r})&\mapsto\Delta^{\text{new}}(\vb{r})=\Delta(\vb{r})e^{\frac{-2i\abs{e}}{\hbar c}\chi(\vb{r})},
    \end{align}
\end{subequations}
where $\chi(\vb{r})$ is an arbitrary scalar function. Owing to the gauge invariance of physical quantities, we can initially choose a convenient gauge condition (or class) to solve the problem. Any other physically equivalent configuration with a different form (regardless of whether it belongs to the same gauge class as the original one or not) can be regarded as the result of a gauge transformation. In what follows, the analysis begins with the Coulomb gauge, $\grad\vdot\vb{A} = 0$, and the gauge class will be progressively restricted through physical arguments and constraints, ultimately leading to complete gauge fixing, in which $\vb{A}(\vb{r})$ is fully determined. 

To diagonalize $H_{\text{eff}}$, we employ the following generalized Bogoliubov transformation~\cite{Halterman_PhysRevB.77.174511}:
\begin{subequations}
    \label{Bogoliubov transformation}
    \begin{align}
        \psi_{\uparrow}(\vb{r})&=\sum\limits_{E_n>0}\left[u_{n\uparrow}(\vb{r})\gamma_n-v^\ast_{n\uparrow}(\vb{r})\gamma^\dagger_n\right],\\
        \psi_{\downarrow}(\vb{r})&=\sum\limits_{E_n>0}\left[u_{n\downarrow}(\vb{r})\gamma_n+v^\ast_{n\downarrow}(\vb{r})\gamma^\dagger_n\right].
    \end{align}
\end{subequations}
The new fermionic operators $\gamma_n$ and $\gamma_n^\dagger$ annihilate and
create Bogoliubov quasiparticles, respectively. Since they diagonalize
$H_{\text{eff}}$, they satisfy the following commutation relations:
\begin{subequations}
    \label{commutator relation}
    \begin{align}
        [H_{\text{eff}},\gamma_n]&=-E_n\gamma_n,\\
        [H_{\text{eff}},\gamma^\dagger_n]&=E_n\gamma^\dagger_n.
    \end{align}
\end{subequations}
Based on Eqs.~\eqref{Bogoliubov transformation} and \eqref{commutator relation}, the evaluation of $\left[\psi_{\sigma}(\vb{r}), H_{\text{eff}}\right]$ leads to the so-called generalized BdG equation:
\begin{subequations}
    \begin{gather}
        \vb{H}_{\text{BdG}}(\vb{r})\vb*{\psi}_n(\vb{r})=E_n\vb*{\psi}_n(\vb{r}),\\
        \vb*{\psi}_n(\vb{r})=\left(u_{n\uparrow}(\vb{r}),u_{n\downarrow}(\vb{r}),v_{n\uparrow}(\vb{r}),v_{n\downarrow}(\vb{r})\right)^{\text{T}},
    \end{gather}
\end{subequations}
where the $4\times4$ matrix $\vb{H}_{\text{BdG}}(\vb{r})$ is given by
\begin{align}
    \vb{H}_{\text{BdG}}(\vb{r})
    &=
    \left(h_{\text{kin}}(\vb{r})\frac{\uptau_0+\uptau_z}{2}-h^{\ast}_{\text{kin}}(\vb{r})\frac{\uptau_0-\uptau_z}{2}\right)\otimes\sigma_0\nonumber\\
    &\ +
    \Big(h_x(y)+\frac{g^*\mu_{B}}{2}[B_{\text{ex},x}+B_{\text{mat},x}(y)]\Big)\uptau_0\otimes\sigma_x\nonumber\\
    &\ +
    \Big(h_z(y)+\frac{g^*\mu_{B}}{2}[B_{\text{ex},z}+B_{\text{mat},z}(y)]\Big)\uptau_z\otimes\sigma_z\nonumber\\
    &\ +
    \left(\Delta(\vb{r})\frac{\uptau_x+i\uptau_y}{2}+\Delta^{\ast}(\vb{r})\frac{\uptau_x-i\uptau_y}{2}\right)\otimes\sigma_{x}.
\end{align}
Here $\uptau_0$ and $\vb*{\uptau}$ denote the identity matrix and Pauli matrices in particle-hole space. The function $h_{\text{kin}}(\vb{r})$ is
\begin{align}
    h_{\text{kin}}(\vb{r})&=\frac{-\hbar^2}{2m^*}\grad^2-E_F-\frac{i\hbar\abs{e}}{m^*c}\vb{A}(\vb{r})\vdot\grad+\frac{e^2\abs{\vb{A}(\vb{r})}^2}{2m^*c^2}.
\end{align}
So far, we have only required $\vb{A}(\vb{r})$ to satisfy the Coulomb gauge condition. However, within the current gauge class, a significant amount of residual freedom remains, as long as $\grad^2\chi(\vb{r}) = 0$ is fulfilled. 


Next, we impose additional physical conditions to further restrict the gauge class.
Based on the structural symmetry of the system, all gauge-invariant local
observables are expected to be independent of $x$ and $z$ and to depend only
on $y$. The corresponding physical translational symmetries need not, in an arbitrary
gauge, be generated by the canonical momentum operators
$-i\hbar\partial_x$ and $-i\hbar\partial_z$, since an ordinary spatial translation may
be accompanied by a gauge transformation. For computational convenience, we
choose a gauge in which both the vector potential and the superconducting
pair potential depend only on $y$,
\begin{equation}
\vb{A}(\vb{r})=\vb{A}(y),
\qquad
\Delta(\vb{r})=\Delta(y).
\end{equation}
In this gauge, translational invariance in the $x$--$z$ plane is manifest,
and the BdG Hamiltonian commutes with $-i\hbar\partial_x$ and
$-i\hbar\partial_z$. Consequently, $k_x$ and $k_z$ are good quantum numbers.


To make the translational symmetry in the $x$--$z$ plane explicit, we further
restrict the Coulomb-gauge class by requiring
\begin{subequations}
    \label{gauge_class_1}
    \begin{align}
        \vb{A}(\vb{r}) &= \vb{A}(y),
        &
        \grad\vdot\vb{A}
        &= \frac{dA_y}{dy}=0,
        \\
        \chi(\vb{r}) &= \chi(y)=a_1y+a_0,
        &
        \grad^2\chi
        &= \frac{d^2\chi}{dy^2}=0.
    \end{align}
\end{subequations}
Within this restricted gauge class, $A_y$ is spatially constant. The residual
gauge transformation shifts this constant according to
$A_y\rightarrow A_y+a_1$ and simultaneously transforms the phase of the pair
potential as $\Delta(y)\rightarrow\Delta(y)\exp\left[\frac{-2i\abs{e}}{\hbar c}\chi(y)\right]$.
Since the allowed gauge functions depend only on $y$, the form
$\Delta(\vb{r})=\Delta(y)$ is preserved within this gauge class. Consequently,
the BdG amplitudes may be chosen in a form that separates the $x$ and $z$
dependences from the remaining $y$ dependence.
Accordingly, $u_{ns}(\vb{r})$ and $v_{ns}(\vb{r})$ may be written as
\begin{subequations}
    \label{solution form}
    \begin{align}
        u_{ns}(\vb{r})&=\frac{1}{\sqrt{\Omega}}e^{i(k_xx+k_zz)}u^{s}_{n}(y),\\
        v_{ns}(\vb{r})&=\frac{1}{\sqrt{\Omega}}e^{i(k_xx+k_zz)}v^{s}_{n}(y).
    \end{align}
\end{subequations}
where $k_x = \frac{2\pi n_x}{L_x}$ and $k_z = \frac{2\pi n_z}{L_z}$, with $n_x, n_z \in \mathbb{Z}$, and $L_x$ and $L_z$ denoting the lengths of the system along the $x$ and $z$ directions, and $\Omega=L_xL_z$. The generalized BdG equation reduces to a quasi-one-dimensional form in space:
\begin{subequations}
    \begin{align}
        &\widetilde{\vb{H}}_{\text{BdG}}(\vb{k}_{\perp},y)\widetilde{\vb*{\psi}}_n(y)
        =
        E_n\widetilde{\vb*{\psi}}_n(y),\\
        &\widetilde{\vb*{\psi}}_n(y)=\left(u_{n}^{\uparrow}(y), u_{n}^{\downarrow}(y), v_{n}^{\uparrow}(y), v_{n}^{\downarrow}(y)\right)^{\text{T}},
    \end{align}
\end{subequations}
where $\vb{k}_\perp\equiv(k_x,0,k_z)$ and $\widetilde{\vb{H}}_{\text{BdG}}(\vb{k}_{\perp},y)$ is given by
\begin{align}
    &\widetilde{\vb{H}}_{\text{BdG}}(\vb{k}_{\perp},y)\nonumber\\
    =&\ 
    h_0(\vb{k}_{\perp},y)\uptau_z\otimes\sigma_0+h_{\grad}(\vb{k}_{\perp},y)\uptau_{0}\otimes\sigma_{0}\nonumber\\
    +&\ 
    \Big(h_x(y)+\frac{g^*\mu_{B}}{2}[B_{\text{ex},x}+B_{\text{mat},x}(y)]\Big)\uptau_0\otimes\sigma_x\nonumber\\
    +&\ 
    \Big(h_z(y)+\frac{g^*\mu_{B}}{2}[B_{\text{ex},z}+B_{\text{mat},z}(y)]\Big)\uptau_z\otimes\sigma_z\nonumber\\
    +&\ 
    \left(\Delta(y)\frac{\uptau_x+i\uptau_y}{2}+\Delta^{\ast}(y)\frac{\uptau_x-i\uptau_y}{2}\right)\otimes\sigma_{x}.
\end{align}
The functions $h_{0}(\vb{k}_{\perp},y)$ and $h_{\grad}(\vb{k}_{\perp},y)$ are:
\begin{subequations}
    \begin{gather}
        h_{0}(\vb{k}_{\perp},y)=E_{\perp}-\frac{\hbar^2}{2m^*}\frac{\partial^2}{\partial y^2}-E_F+\frac{\abs{e}^2\abs{\vb{A}(y)}^2}{2m^*c^2},\\
        h_{\grad}(\vb{k}_{\perp},y)=\frac{\hbar\abs{e}}{m^*c}\vb{k}_{\perp}\vdot\vb{A}(y)-\frac{i\hbar\abs{e}}{m^*c}C_{A_y}\frac{\partial}{\partial y},
    \end{gather}
\end{subequations}
where $E_\perp=\frac{\hbar^2\abs{\vb{k}_\perp}^2}{2m^*}$ 
and we denote $A_y(y)$ to be a constant $C_{A_y}$ here.
The eigenstates are now labeled by both the in-plane momentum
$\vb{k}_{\perp}\equiv(k_x,0,k_z)$ and the band index $n$. Accordingly, the
sum over all BdG eigenstates can be written as $\sum_{E_n}\equiv\sum_{\vb{k}_{\perp}}\sum_{E_n(\vb{k}_{\perp})}$,
where the inner sum runs over the eigenstates of
$\widetilde{\vb H}_{\mathrm{BdG}}(\vb{k}_{\perp},y)$ for each fixed
$\vb{k}_{\perp}$.

Using Eqs.~\eqref{Bogoliubov transformation} and \eqref{solution form}, we now
verify that the assumption $\Delta(\mathbf r)=\Delta(y)$ is preserved by the
self-consistency equation:
\begin{align}
    \Delta(\mathbf r)
    &=
    \mathcal{V}_\text{eff}(y)\langle\psi_{\uparrow}(\mathbf r)\psi_{\downarrow}(\mathbf r)\rangle
    \nonumber\\
    &=
    \frac{\mathcal{V}_\text{eff}(y)}{2\Omega}\sum\limits_{E_n}{\vphantom{\sum}}'
    \left[
        u^{\uparrow}_{n}(y)v^{\downarrow\ast}_{n}(y)
        +u^{\downarrow}_{n}(y)v^{\uparrow\ast}_{n}(y)
    \right]
    \tanh\left(\frac{E_n}{2k_BT}\right)
    \nonumber\\
    &=
    \Delta(y).
    \label{Delta potential}
\end{align}
Here,
$\sum^{\prime}_{E_n}
\equiv\sum_{0<E_n\leq\hbar\omega_D}$,
where $\hbar\omega_D$ denotes the pairing-interaction energy cutoff.
Equation~\eqref{Delta potential} therefore confirms that the form
$\Delta(\mathbf r)=\Delta(y)$ is self-consistently preserved. The associated Cooper pair amplitude $F_{\Delta}(\vb{r})=F_{\Delta}(y)$ can now be written through the relation $\Delta(y)=\mathcal{V}_\text{eff}(y)F_{\Delta}(y)$. 
$F_{\Delta}$ is normalized to $\Delta_{\text{bulk}}/\mathcal{V}_\text{eff}$ in our work. Next, we discuss the boundary conditions for $u_n^s(y)$ and $v_n^s(y)$.
Since the electrons are confined within the heterostructure, we impose open
boundary conditions,
\begin{equation}
u_n^s(y=0)=v_n^s(y=0)=u_n^s(y=d)=v_n^s(y=d)=0,
\end{equation}
where $d=d_{\mathrm{F1}}+d_{\mathrm{F2}}+d_{\mathrm S}$. 
Accordingly, the quasiparticle amplitudes can be expanded in a sine basis as
\begin{subequations}
\begin{align}
u_n^s(y)
&=
\sqrt{\frac{2}{d}}
\sum_{a=1}^{N}
u_{na}^s
\sin\left(\frac{a\pi y}{d}\right),
\\
v_n^s(y)
&=
\sqrt{\frac{2}{d}}
\sum_{a=1}^{N}
v_{na}^s
\sin\left(\frac{a\pi y}{d}\right).
\end{align}
\end{subequations}
Here, $N$ denotes the numerical cutoff for the basis expansion. This allows us to further rewrite the BdG equation in the following form:
\begin{subequations}
    \begin{gather}
        \vb*{\mathcal{H}}_{\text{BdG}}\vb*{\Psi}_n=\varepsilon_n\vb*{\Psi}_n\text{ with }\varepsilon_n=\frac{E_n}{E_F},\\
        \vb*{\Psi}_n=[u_{na=1\sim N}^{\uparrow},u_{na=1\sim N}^{\downarrow},v_{na=1\sim N}^{\uparrow},v_{na=1\sim N}^{\downarrow}]^{\text{T}},
    \end{gather}
\end{subequations}
where the $4N\times4N$ matrix $\vb*{\mathcal{H}}_{\text{BdG}}(\vb{k}_{\perp})$ is given by:
\begin{widetext}
\begin{align}
    \vb*{\mathcal{H}}_{\text{BdG}}(\vb{k}_{\perp})=
    \begin{bmatrix}
        (\vb*{\mathcal{H}}_{+}+\vb*{\mathcal{H}}_{z}+\vb*{\mathcal{B}}_z)&\vb*{\mathcal{H}}_x+\vb*{\mathcal{B}}_x&\vb*{0}&\vb*{\mathcal{D}}\\
        \vb*{\mathcal{H}}_x+\vb*{\mathcal{B}}_x&(\vb*{\mathcal{H}}_{+}-\vb*{\mathcal{H}}_{z}-\vb*{\mathcal{B}}_z)&\vb*{\mathcal{D}}&\vb{0}\\
        \vb{0}&\vb*{\mathcal{D}}^{\ast}&-(\vb*{\mathcal{H}}_{-}+\vb*{\mathcal{H}}_{z}+\vb*{\mathcal{B}}_z)&\vb*{\mathcal{H}}_x+\vb*{\mathcal{B}}_x\\
        \vb*{\mathcal{D}}^{\ast}&\vb{0}&\vb*{\mathcal{H}}_x+\vb*{\mathcal{B}}_x&-(\vb*{\mathcal{H}}_{-}-\vb*{\mathcal{H}}_{z}-\vb*{\mathcal{B}}_z)\label{BdG_mode_first}
    \end{bmatrix}.
\end{align}
\end{widetext}
\clearpage
The quantities
$\vb*{\mathcal{H}}_{\pm}
=\vb*{\mathcal{H}}_{0}\pm\vb*{\mathcal{H}}_{\grad}$,
$\vb*{\mathcal{H}}_{z}$,
$\vb*{\mathcal{H}}_{x}$,
$\vb*{\mathcal{B}}_{z}$,
$\vb*{\mathcal{B}}_{x}$, and
$\vb*{\mathcal{D}}$
are all $N\times N$ matrices. Their matrix elements, expressed in
dimensionless form, are given by
\begin{widetext}
\begin{subequations}
    \begin{align}
        (\vb*{\mathcal{H}}_0)_{ab}
        &=
        \left[\frac{E_{\perp}}{E_F}+\left(\frac{a\pi}{k_Fd}\right)^2-1\right]\delta_{ab}
        +
        \frac{2}{d}\left(\frac{\abs{e}}{\hbar k_Fc}\right)^2\int_{0}^{d}\vb{A}^2(y)
        \left[
            \mathcal{J}_{a}(y)\mathcal{J}_{b}(y)
        \right]dy,
        \\
        (\vb*{\mathcal{H}}_{\grad})_{ab}
        &=
        \frac{4\abs{e}}{dc\hbar k_F^2}\int_{0}^{d}\left[k_xA_x(y)+k_zA_z(y)\right]
        \left[
            \mathcal{J}_{a}(y)\mathcal{J}_{b}(y)
        \right]dy
        \\
        (\vb*{\mathcal{H}}_z)_{ab}
        &=
        \frac{1}{E_F}
        \bigg\{
            (h_1-h_2\cos{\theta})\Big[\mathcal{K}_{a-b}(d_{F_1})-\mathcal{K}_{a+b}(d_{F_1})\Big]\nonumber
        \\
        &\qquad\qquad\ \ +
            h_2\cos{\theta}\Big[\mathcal{K}_{a-b}(d_{F_1}+d_{F_2})-\mathcal{K}_{a+b}(d_{F_1}+d_{F_2})\Big]
        \bigg\}\text{ for }a\neq b,
        \\
        &=
        \frac{1}{E_F}
        \bigg\{
            \frac{h_1d_{F_1}+h_2(\cos{\theta})d_{F_2}}{d}-(h_1-h_2\cos{\theta})\mathcal{K}_{2a}(d_{F_1})-h_2(\cos{\theta})\mathcal{K}_{2a}(d_{F_1}+d_{F_2})
        \bigg\}\text{ for }a=b,
        \\
        (\vb*{\mathcal{H}}_x)_{ab}
        &
        =
        \frac{1}{E_F}
        \bigg\{
            -h_2\sin{\theta}\Big[\mathcal{K}_{a-b}(d_{F_1})-\mathcal{K}_{a+b}(d_{F_1})\Big]\nonumber
        \\
        &\qquad\qquad\quad
            +h_2\sin{\theta}\Big[\mathcal{K}_{a-b}(d_{F_1}+d_{F_2})-\mathcal{K}_{a+b}(d_{F_1}+d_{F_2})\Big]
        \bigg\}\text{ for }a\neq b,
        \\
        &
        =
        \frac{1}{E_F}
        \bigg\{
            \frac{h_2(\sin{\theta})d_{F_2}}{d}+h_2(\sin{\theta})\mathcal{K}_{2a}(d_{F_1})-h_2(\sin{\theta})\mathcal{K}_{2a}(d_{F_1}+d_{F_2})
        \bigg\}\text{ for }a=b,
        \\
        (\vb*{\mathcal{B}}_{z})_{ab}
        &=
        \frac{g^*}{2}\frac{\mu_B B_{\text{ex},z}}{E_F}\delta_{ab}+\frac{g^*}{2}\frac{2\mu_B}{E_Fd}\int^{d}_{0}\mathcal{J}_a(y)B_{\text{mat},z}(y)\mathcal{J}_{b}(y)dy,
        \\
        (\vb*{\mathcal{B}}_{x})_{ab}
        &=
        \frac{g^*}{2}\frac{\mu_B B_{\text{ex},x}}{E_F}\delta_{ab}+\frac{g^*}{2}\frac{2\mu_B}{E_Fd}\int^{d}_{0}\mathcal{J}_a(y)B_{\text{mat},x}(y)\mathcal{J}_{b}(y)dy,
        \\
        (\vb*{\mathcal{D}})_{ab}
        &
        =
        \frac{2}{E_Fd}
        \int_{d_{F_1}+d_{F_2}}^{d_{F_1}+d_{F_2}+d_S}\mathcal{J}_a(y)\Delta(y)\mathcal{J}_b(y)dy.\label{D}
    \end{align}
\end{subequations}
\end{widetext}
Here the functions $\mathcal{J}_a(y)$ and $\mathcal{K}_a(y)$ are defined by
    \begin{align*}
        \mathcal{K}_a(y)\equiv\frac{\sin(\frac{a\pi y}{d})}{a\pi}\equiv\frac{\mathcal{J}_a(y)}{a\pi}.
    \end{align*}
Before proceeding to the most intricate part—the self-consistent calculation
of the vector potential—we first define several physical quantities that act
as sources of the electromagnetic field and thereby enter the determination
of the vector potential through Maxwell's equations. The first quantity is the total current, $\vb{J}(\vb{r})=\frac{-i\hbar\abs{e}}{2m^*}\sum\limits_{s}\langle[\grad\psi^\dagger_s(\vb{r})]\psi_s(\vb{r})-\psi^\dagger_s(\vb{r})[\grad\psi_s(\vb{r})]\rangle-\frac{\abs{e}^2}{m^*c}\sum\limits_s\vb{A}(\vb{r})\langle\psi^\dagger_s(\vb{r})\psi_s(\vb{r})\rangle$. In our geometry, its components along each direction are given by
\begin{widetext}
\begin{subequations}
    \begin{align}
        J_i(y)
        &=
        \frac{-\hbar\abs{e}}{m^*\Omega}\sum_{\vb{k}_\perp}\sum_{E_{n(\vb{k}_\perp)}>0}
        \Big(k_i
        \sum_s\abs{u^{s}_{n}(y)}^2n_F(E_n)
        -
        k_i
        \sum_s\abs{v^{s}_{n}(y)}^2\big[1-n_F(E_n)\big]
        \Big)\nonumber
        \\
        &\quad-
        \frac{\abs{e}^2}{m^*c\Omega}A_i(y)\sum_{E_n>0}
        \Big(
        \sum_{s}\abs{u^{s}_{n}(y)}^2n_F(E_n)
        +
        \sum_{s}\abs{v^{s}_{n}(y)}^2
        \big[1-n_F(E_n)\big]
        \Big)\nonumber
        \\
        &=
        J^{P}_i(y)+J^{D}_i(y)=J_i(y)\text{ for }i=x\text{ or }z\ ,\label{J_i}
        \\
        \nonumber
        \\
        J_{y}(y)
        &=
        \frac{\hbar\abs{e}}{m^*\Omega}\sum_{E_n>0}
        \Big(
        \sum_s\Im\big[u^{s}_{n}(y)\frac{\partial}{\partial y}u_{n}^{s\ast}(y)\big]n_F(E_n)
        +
        \sum_s\Im\big[v^{s\ast}_{n}(y)\frac{\partial}{\partial y}v^{s}_{n}(y)\big]\big[1-n_F(E_n)\big]
        \Big)\nonumber
        \\
        &\quad-
        \frac{\abs{e}^2}{m^*c\Omega}C_{A_y}\sum_{E_n>0}
        \Big(
        \sum_{s}\abs{u^{s}_{n}(y)}^2n_F(E_n)
        +
        \sum_{s}\abs{v^{s}_{n}(y)}^2\big[1-n_F(E_n)\big]
        \Big)\nonumber
        \\
        &=
        J^{P}_y(y)+J^{D}_y(y)=J_y(y).\label{J_y}
    \end{align}
\end{subequations}
\end{widetext}
It is convenient to normalize these current components by $J_0=\frac{k_F^3}{3\pi^2}\abs{e}v_F$,
where $v_F$ is the Fermi velocity. Another relevant physical quantity is the
local magnetization,
\begin{equation}
\vb{M}(\vb{r})
=
-\frac{g^*}{2}\mu_B
\sum_{ss'}
\left\langle
\psi_s^\dagger(\vb{r})
\vb*{\sigma}_{ss'}
\psi_{s'}(\vb{r})
\right\rangle.
\end{equation}
For the geometry considered here, its components can be written as
\begin{widetext}
\begin{subequations}
    \begin{align}
        M_x(y)
        &=
        -\frac{g^*}{2}\frac{2\mu_B}{\Omega}\sum_{E_n>0}
        \Big(
        \Re\big[u_{n}^{\uparrow\ast}(y)u_{n}^{\downarrow}(y)\big]n_F(E_n)
        -
        \Re\big[v_{n}^{\uparrow}(y)v_{n}^{\downarrow\ast}(y)\big]\big[1-n_F(E_n)\big]
        \Big)
        \\
        \nonumber
        \\
        M_z(y)
        &=
        -\frac{g^*}{2}\frac{\mu_B}{\Omega}\sum_{E_n>0}
        \Big(
        \Big[\abs{u^{\uparrow}_{n}(y)}^2-\abs{u^{\downarrow}_{n}(y)}^2\Big]n_F(E_n)
        +
        \Big[\abs{v^{\uparrow}_{n}(y)}^2-\abs{v^{\downarrow}_{n}(y)}^2\Big]
        \big[1-n_F(E_n)\big]
        \Big)
        \\
        \nonumber
        \\
        M_y(y)
        &=
        -\frac{g^*}{2}\frac{2\mu_B}{\Omega}\sum_{E_n>0}
        \Big(
        \Im\big[u_{n}^{\uparrow\ast}(y)u^{\downarrow}_{n}(y)\big]n_F(E_n)
        -
        \Im\big[v^{\uparrow}_{n}(y)v_{n}^{\downarrow\ast}(y)\big]\big[1-n_F(E_n)\big]
        \Big).\label{M_y}
    \end{align}
\end{subequations}
\end{widetext}
For convenience, we normalize these components by $M_0=\frac{k_F^3}{3\pi^2}\mu_B$.\par

Let us now proceed to the most technically demanding part of the formulation, namely, determining the vector potential self-consistently. To accomplish this task, we employ Maxwell's equations. Since the system is in equilibrium, it suffices to consider only the Amp\`ere-Maxwell law, i.e., $\grad\times\vb{B}(\vb{r})=\frac{4\pi}{c}\left[\vb{J}(\vb{r})+c\grad\cross\vb{M}(\vb{r})\right]$. In our gauge choice, this relation can be rewritten as follows:
\begin{subequations}
    \begin{gather}
        \frac{\partial^2}{\partial y^2}A_{x}(y)=-4\pi\left[\frac{J_x(y)}{c}+\frac{\partial}{\partial y}M_z(y)\right],\label{Maxwell's Eqs_x}\\
        \frac{\partial^2}{\partial y^2}A_{z}(y)=-4\pi\left[\frac{J_z(y)}{c}-\frac{\partial}{\partial y}M_x(y)\right],\label{Maxwell's Eqs_z}\\
        \frac{\partial^2}{\partial y^2}C_{A_y}=-\frac{4\pi}{c}J_{y}(y)=0.\label{J_y=0}
    \end{gather}
\end{subequations}
Because $\vb{A}$ and $\vb{M}$ are functions of $y$ only and $A_y=C_{A_y}$, 
we have $(\nabla\times\vb{B})_y=c(\nabla\times\vb{M})_y=0$ which leads to Eq.~\eqref{J_y=0}.
Since $C_{A_y}$ represents a
residual gauge degree of freedom, it may be set to any convenient constant.
This freedom arises from the $a_1$ term in
$\chi(y)=a_1y+a_0$, which shifts $A_y$ by a constant and simultaneously
changes the spatial phase of the superconducting pair potential. By contrast,
the $a_0$ term leaves $\vb{A}(\vb{r})$ unchanged and produces only a global
phase transformation of the fermionic fields and the pair potential.

Physically, Eq.~\eqref{J_y=0} means the
solution is in the sector with vanishing gauge-invariant condensate momentum
along $y$. 
This reflects the fact that the system is in equilibrium without 
an imposed phase bias or current along the $y$ direction. 
The residual gauge freedom may then be used to set $A_y=0$, which
in this sector implies that the phase of $\Delta(y)$ is spatially uniform.
The remaining global phase freedom can subsequently be used to choose
$\Delta(y)$ to be real.
By choosing $\Delta(y) \in \mathbb{R}$, the matrix $\vb*{\mathcal{D}}$ in Eq.~\eqref{D}, becomes an $N\times N$ real matrix, which in turn makes the BdG Matrix, Eq.~\eqref{BdG_mode_first}, a real symmetric $4N\times 4N$ matrix.
Consequently, its eigenvectors may be chosen to be real,
\begin{equation}
\boldsymbol{\Psi}_n\in\mathbb{R}^{4N}
\quad\Longrightarrow\quad
u_n^s(y),\,v_n^s(y)\in\mathbb{R}.
\label{real_property}
\end{equation}
Substituting these real amplitudes into the self-consistency equation for the
pair potential yields a real $\Delta(y)$. 
Moreover, from Eq.~\eqref{M_y}, it is easy to see
\begin{equation}
M_y(y)=0,
\end{equation}
as expected from the setup of our system.
At the same time, with these real amplitudes and the fact that $A_y=0$,
Eq.~\eqref{J_y} now manifestly shows 
\begin{equation}
J_y(y)=0,
\end{equation}
consistent with Eq.~\eqref{J_y=0}.
Thus, the self-consistent iteration preserves the conditions
$A_y=0$ and $\Delta(y)\in\mathbb{R}$, confirming the internal consistency of
the chosen gauge-fixed formulation.

To preserve the pair potential exactly, the remaining constant gauge
transformation must satisfy
\begin{equation}
\exp\left(-\frac{2i|e|}{\hbar c}\chi\right)=1.
\end{equation}
Consequently, the allowed residual gauge functions are restricted to
\begin{equation}
\chi=n\frac{\pi\hbar c}{|e|},
\qquad n\in\mathbb Z.
\end{equation}
These transformations leave both $\mathbf A$ and $\Delta$ unchanged and
therefore represent a physically trivial residual gauge redundancy.
Within the restricted gauge class $\chi(\mathbf r)=\chi(y)$, the residual
gauge transformation affects only $A_y(y)$, while $A_x(y)$ and $A_z(y)$
remain unchanged. Therefore, after fixing $A_y(y)=0$, the vector potential is
uniquely specified within this gauge class, up to the remaining discrete
global $\mathbb Z_2$ redundancy. The resulting gauge-fixed formulation can be
summarized as
\begin{subequations}
    \label{final_gauge}
    \begin{gather}
        A_{x}(y),A_{z}(y)\text{ are unique and }A_y(y)=0\,;\\
        \Delta(\vb{r})=\Delta(y)\in\mathbb{R}\text{ and }\chi=0,\pm\frac{\pi\hbar c}{\abs{e}},\pm\frac{2\pi\hbar c}{\abs{e}},\cdots;\\
        \vb*{\mathcal{H}}_{\text{BdG}}(\vb{k}_{\perp})\vb*{\Psi}_n=\varepsilon_n\vb*{\Psi}_n\text{ with }\vb*{\Psi}_n\in\mathbb{R}^{4N}.
    \end{gather}
\end{subequations}
Finally, using the formulation, Eqs.~\eqref{final_gauge}, together with Eqs.~\eqref{Maxwell's Eqs_x} and \eqref{Maxwell's Eqs_z}, the self-consistent expressions for $A_x(y)$ and $A_z(y)$ can be obtained. Equations~\eqref{Maxwell's Eqs_x} and ~\eqref{Maxwell's Eqs_z} are standard inhomogeneous second-order ODEs, with $-4\pi\left[\frac{1}{c}J_x(y)+\frac{\partial}{\partial y}M_z(y)\right]$ and $-4\pi\left[\frac{1}{c}J_z(y)-\frac{\partial}{\partial y}M_x(y)\right]$ serving as the source terms. The Green's function $G(y, y')$ corresponding to this type of ODE is given by:
\begin{align}
    G(y,y^\prime)\equiv\frac{1}{2}\abs{y-y^\prime}.
\end{align}
It is not difficult to check that $f(y)=\int^{\infty}_{-\infty}G(y,y^\prime)g(y^\prime)dy^\prime$ satisfies $\frac{d^2}{dy^2}f(y)=g(y).$ As a result, we write down the following expressions for $A_x(y)$ and $A_z(y)$:
\begin{widetext}
\begin{subequations}
    \label{vector potential}
    \begin{align}
        A_{x}(y)&=
        -2\pi\int^{d}_{0}
        \left[
        \abs{y-y^\prime}\frac{J_x(y^\prime)}{c}+\frac{y-y^\prime}{\abs{y-y^\prime}}M_z(y^\prime)
        \right]dy^\prime
        -B_{\text{ex},z}\,y+A_{cx}\,,
        \\
        A_{z}(y)&=
        -2\pi\int^{d}_{0}
        \left[
        \abs{y-y^\prime}\frac{J_z(y^\prime)}{c}-\frac{y-y^\prime}{\abs{y-y^\prime}}M_x(y^\prime)
        \right]dy^\prime
        +B_{\text{ex},x}\,y+A_{cz}\,.
    \end{align}
\end{subequations}
The corresponding total magnetic field, $\mathbf B(y)=\nabla\times\mathbf A(y)$, is
\begin{subequations}
    \label{B fields}
    \begin{align}
        B_x(y)
        &=
        \frac{\partial}{\partial y}A_{z}(y)
        =
        -2\pi\int^{d}_{0}\frac{y-y^\prime}{\abs{y-y^\prime}}\frac{J_z(y^\prime)}{c}dy^\prime
        +4\pi M_{x}(y)
        +B_{\text{ex},x}\nonumber
        \\
        &=
        B_{J,x}(y)+B_{M,x}(y)+B_{\text{ex},x}=B_{\text{mat},x}(y)+B_{\text{ex},x}\,,
        \\
        B_z(y)
        &=
        -\frac{\partial}{\partial y}A_{x}(y)
        =
        2\pi\int^{d}_{0}\frac{y-y^\prime}{\abs{y-y^\prime}}\frac{J_x(y^\prime)}{c}dy^\prime
        +4\pi M_{z}(y)
        +B_{\text{ex},z}\nonumber
        \\
        &=
        B_{J,z}(y)+B_{M,z}(y)+B_{\text{ex},z}=B_{\text{mat},z}(y)+B_{\text{ex},z}.
    \end{align}
\end{subequations}
\end{widetext}
It is worth noting that the integration constants $A_{cx}$ and $A_{cz}$ appearing in Eqs.~\eqref{vector potential} are not arbitrary; they will be determined by the conditions specified below. From Eqs.~\eqref{B fields}, we can observe that $B_{(x,z)}(y > d)=B_{(x,z)}(d)$ and $B_{(x,z)}(y < 0)=B_{(x,z)}(0).$ Physically, we may expect that at distances far from the system, the magnetic field should approach that of the externally applied uniform field. In other words, $B_{x}(0)=B_{x}(d)=B_{\text{ex},x}$ and $B_{z}(0)=B_{z}(d)=B_{\text{ex},z}$. For clarity and conciseness, these conditions amount to the following compact form:
\begin{align}
    \int^{d}_{0}J_{i}(y)\,dy=0\text{ for }i=x,z.\label{current_flux_conservation}
\end{align}
This condition expresses the physically intuitive requirement that, in the absence of an external voltage and a conducting loop, the net current flux across the $x$-$y$ and $y$-$z$ planes must vanish. The constants $A_{cx}$ and $A_{cz}$, which are to be determined self-consistently, are updated at the end of each iteration using Eq.~\eqref{current_flux_conservation}. Therefore, the expressions for all observable physical quantities and self-consistently updated variables in the system have been fully established.

In summary, the coupled BdG--Maxwell formulation developed above determines 
the superconducting pair potential, quasiparticle wavefunctions, current density,
magnetization, and vector potential self-consistently. The electromagnetic
boundary conditions further impose vanishing net in-plane sheet currents,
which fix the remaining constants in the vector potential. These equations
provide the basis for the numerical calculations presented in the following
section.\par

Finally, since the subsequent numerical results involve the odd-frequency triplet correlations, it is useful to specify their time dependence. Because the system considered here is in equilibrium and therefore possesses time-translational invariance, these correlations depend only on the relative time $t-t^\prime$. Without loss of generality, $t^\prime$ can thus be set to zero. Furthermore, owing to their odd-frequency symmetry, these correlations are odd when the relative time is reversed. It is sufficient to present the expressions for $t>0$, as given below:
\begin{widetext}
\begin{subequations}
    \begin{align}
        f_0(y,t>0,t^\prime=0)
        &=
        \frac{1}{2\Omega}\sum_{E_n>0}\left[ u^{\uparrow}_n(y)v^{\downarrow *}_n(y)-u^{\downarrow}_n(y)v^{\uparrow *}_n(y)\right]\left[\cos(\frac{E_n}{\hbar}t)-i\sin(\frac{E_n}{\hbar}t)\tanh(\frac{E_n}{2k_BT})\right],\label{triplet correlations_a}
        \\
        f_{\pm}(y,t>0,t^\prime=0)
        &=
        \frac{-1}{2\Omega}\sum_{E_n>0}\left[u^{\uparrow}_n(y)v^{\uparrow *}_n(y)\mp u^{\downarrow}_n(y)v^{\downarrow *}_n(y)\right]\left[\cos(\frac{E_n}{\hbar}t)-i\sin(\frac{E_n}{\hbar}t)\tanh(\frac{E_n}{2k_BT})\right].\label{triplet correlations_b}
    \end{align}
\end{subequations}
\end{widetext}

\section{Results}\label{Results}
In this section, we present the results of the self-consistent calculations
based on the coupled BdG--Maxwell framework introduced in Sec.~\ref{methods}.
Our numerical self-consistency criterion requires that
the relative changes between two successive iterations, i.e.,
\begin{align*}
    \epsilon_{\text{rel}}=\frac{\displaystyle\int^{d}_{0}dy\, \left\|\vb{X}^{(n)}(y)-\vb{X}^{(n-1)}(y)\right\|}{\displaystyle\int^{d}_{0}dy\, \left\|\vb{X}^{(n)}(y)\right\|},
\end{align*}
be less than $5\times 10^{-4}$ for both the pair amplitudes and the vector potentials, 
where $\vb{X}^{(n)}(y)$ denotes the corresponding quantity at the \(n\)th iteration, 
and $\|\cdot\|$ denotes the absolute value for a scalar quantity and the Euclidean norm for a vector quantity.
In evaluating quantities involving the summation $\sum_{\vb{k}_{\perp}}$, 
we uniformly partition the square momentum-space domain into $1200\times1200$ cells 
and use the midpoint of each cell as the representative $\vb{k}_{\perp}$ point.
Before proceeding to the detailed discussion, it is necessary to clarify 
the intrinsic numerical uncertainties and limitations associated with our approach. 
There are two main sources of numerical uncertainty in the normalized current density results.
First, the electric current calculation 
involves summing over the transverse momentum $\vb{k}_{\perp}$ on a discretized 
two-dimensional $\vb{k}_{\perp}$ grid. This introduces a systematic discretization error associated with the finite
number of momentum points used in the calculation. Under the dimensionless units adopted in this work, the residual numerical
uncertainty in the current density is of order $10^{-6}J_0$.
This maximum grid error is estimated using a free-electron gas with a nonzero constant transverse vector potential, i.e., $A_y=0$. Its magnitude is also comparable to that found in the collinear cases $(\vb{h}_1\parallel\vb{h}_2\parallel\hat{\vb{z}})$, where $J_z$ is expected to vanish.
Accordingly,
spontaneous current-density values comparable to or smaller than this scale should be
interpreted with caution and are not considered physically significant.
This is also why we perform the calculations at a finite but low temperature.
The finite-temperature Fermi distribution smooths the contributions from
individual $\vb{k}_{\perp}$ points and thereby reduces the discretization error
associated with the finite momentum-space grid. In practice, this amounts to
using the smooth occupation factor $n_F(E_n)$ instead of the zero-temperature
step function $\theta(-E_n)$.

In addition to the grid error discussed above, 
we also observe current-density uncertainties arising from different initial profiles for $\Delta(y)$. 
Since the systems examined in this work are primarily considered in the absence of 
external magnetic fields, $\vb{A}(y)$ is initially set to zero. 
Consequently, the convergence of $\vb{A}(y)$ during the early stages of iteration 
may be affected by different initial uniform guesses for $\Delta(y)$. 
Although $\Delta(y)$ and $\vb{M}(y)$ are highly robust quantities and remain essentially 
unaffected by variations in the initial guess of $\Delta(y)$, the quantities 
$\vb{A}(y)$ and $\vb{J}(y)$ are comparatively more sensitive and are explicitly 
coupled to each other. Such sensitivity may cause the self-consistent iterations 
to gradually deviate along different convergence trajectories of $\vb{A}(y)$, 
leading to appreciable deviations in the converged normalized current density $\vb{J}(y)/J_0$. 
Based on our observations, such differences in $\vb{J}(y)/J_0$
are usually negligible and at most $\lesssim10^{-5}$. 
For all results presented in this paper this
initialization-dependent uncertainty is $\lesssim10^{-5}$.
Nevertheless, we note that this uncertainty may become appreciable ($\sim 10^{-4}$) in 
rare situations, i.e., 
when the exchange fields in the two ferromagnetic layers
are both very weak and essentially equal in magnitude, $h_1\approx h_2\approx 0.01 E_F$. 
Importantly, this variation does not alter
any of the qualitative conclusions of this work.
In the results presented in this work, 
we first carry out a calculation with electromagnetic effects included for an arbitrary 
uniform guess of $\Delta(y)$
to obtain a stable self-consistent $\Delta(y)$ 
whenever the variation in the converged current density
arising from different initial guesses cannot be neglected,
i.e., when it is $\gtrsim 10^{-6}$.
We then use this $\Delta(y)$ again as the initial profile for a 
second self-consistent calculation. 
We expect that this procedure effectively reduces the influence of variations 
in the initial guess of $\Delta(y)$ on the final results.\par

As noted in the preceding paragraph, different initial profiles of $\Delta(y)$ can all lead to converged solutions. Although the converged $\Delta(y)$ and magnetization $\vb{M}(y)$ remain essentially unchanged, the resulting spontaneous currents $\vb{J}(y)$ can differ. This behavior suggests the existence of multiple self-consistent branches associated with essentially the same converged $\Delta(y)$.
The appearance of such multivalued solutions is not unexpected in the present problem. No additional approximation or restrictive constraint, such as imposing a London relation between $\vb{A}(y)$ and $\vb{J}(y)$, is introduced; instead, the Maxwell and BdG equations are solved fully self-consistently. Consequently, determining $\Delta(y)$ and $\vb{A}(y)$ amounts to repeatedly solving a highly nonlinear coupled system, for which multiple self-consistent solutions with some dependence on the initial conditions may arise. 
That is, the fixed-point iteration can
converge to distinct self-consistent solutions depending on the starting
point.

In the discussion below, we rescale all lengths by the Fermi wave vector $k_F$, defining $Y=k_Fy$. 
Unless otherwise specified, we fix the layer thicknesses to $k_Fd_{F1}=50$, $k_Fd_{F2}=10$, and $k_Fd_S=75$, 
and take the superconducting coherence length to be $k_F\xi=50$. For qualitative
analysis and computational convenience, the remaining parameters are chosen
as $m^*=m_e$, $g^*=2$, $\hbar\omega_D=0.04E_F$,
$\Delta_{\mathrm{bulk}}=2E_F/(\pi k_F\xi)$,
and $k_BT=5\times10^{-4}E_F$.
Although the BdG equations are expressed in
dimensionless form, the electromagnetic coupling in the coupled
BdG--Maxwell formulation depends on the ratio $v_F/c$. We therefore take
$v_F/c=1/300$, corresponding to $v_F\simeq10^8\,\mathrm{cm/s}$.

\begin{figure*}[ht!]
    \includegraphics[width=\textwidth,clip]{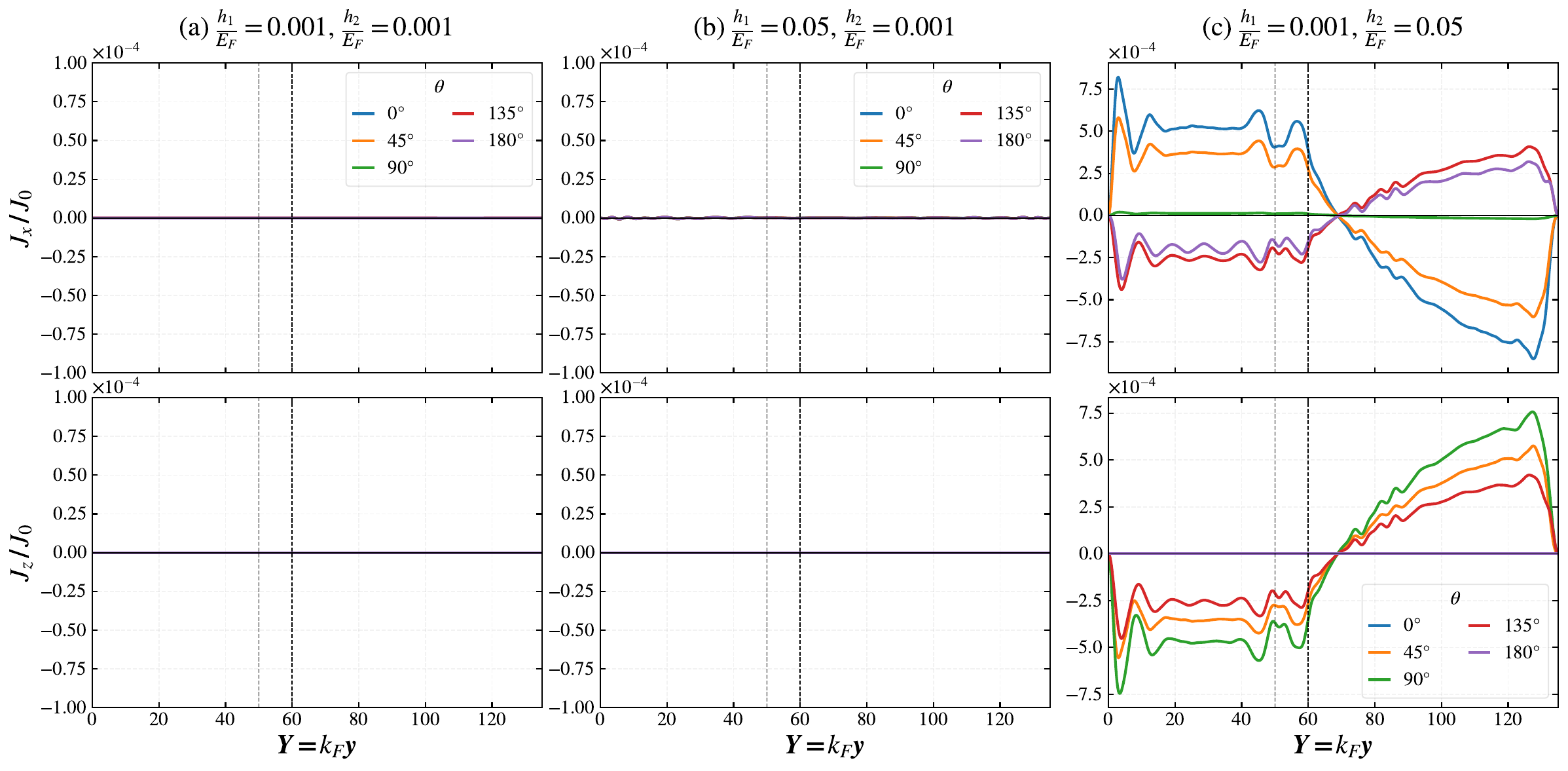}
    \caption{Spatial distributions of the spontaneous current density in the heterostructure for three weak-exchange-field configurations, demonstrating the switch-like role of the F2 layer in generating spontaneous currents. Except for the exchange-field parameters, all other parameters are fixed at the common values specified in the main text. The vertical black dashed lines indicate the interfaces between adjacent layers.}
    \label{switch_effect_weak_case}
\end{figure*}
\subsection{Regime of weak exchange field}
We first discuss an important switching effect due to the electromagnetic coupling. 
Similar to the conventional superconducting proximity effects~\cite{Wu_PhysRevB.86.014523}, 
we find that, compared with the F1 layer, the F2 layer plays a dominant role 
in the generation of spontaneous currents, i.e. its exchange interaction strength 
strongly influences whether an appreciable current develops in the system. 
To illustrate this behavior, we consider two representative regimes.
For the first regime, the exchange fields 
in the two ferromagnetic layers are both weak  (see Fig.~\ref{switch_effect_weak_case}).
In this regime, spontaneous currents emerge only when the exchange field in the F2 layer is increased, 
whereas increasing $h_1$ alone does not produce an appreciable spontaneous current.
From Fig.~\ref{switch_effect_weak_case}(a), we see  that, when the exchange fields 
in both F layers are very weak ($h_1=h_2=0.001E_F$), 
no appreciable spontaneous current is generated at any
relative angle $\theta$.
This behavior is consistent with our expectation, as the exchange fields act as 
the initial driving sources in the heterostructure, the electromagnetic proximity effects 
can be neglected with this small strength of exchange interaction. 

In Fig.~\ref{switch_effect_weak_case}, we further examine cases in which only
one of the two exchange-field strengths is varied.
Figure~\ref{switch_effect_weak_case}(b) clearly shows that increasing $h_1$
while keeping $h_2$ fixed has almost no effect. In contrast,
Fig.~\ref{switch_effect_weak_case}(c) shows that increasing $h_2$ while
keeping $h_1$ fixed induces pronounced spontaneous current densities.
Furthermore, the resulting current-density profiles exhibit switch-like
behavior. This can be regarded as an important manifestation of the electromagnetic
proximity effect. An enhanced exchange field in the F2 layer adjacent to the
superconductor induces a current in the superconducting layer, which is then
redistributed throughout the heterostructure through the current-conservation
condition in Eq.~\eqref{current_flux_conservation}. In contrast,
because the F1 layer is separated from S by F2, enhancing $h_1$
has only a weak influence on the electromagnetic response in the
superconducting layer.

The angular dependence of the current-density profiles in Fig.~\ref{switch_effect_weak_case}(c) provides further support for the switch-like effect of the F2 layer, indicating that the spontaneous-current response is governed primarily by $h_2$ when $h_1 \ll h_2$. According to the Biot--Savart law, the magnetic field associated with a given current component is transverse to the current direction. 
Accordingly, in the present geometry, the self-consistent electromagnetic
coupling connects $J_x$ with the $z$-directed magnetic-field component $B_z$,
whereas $J_z$ is coupled to the $x$-directed component $B_x$.
When $h_1\ll h_2$, the transverse component of the exchange field in F2
provides the primary magnetic driving that determines the emergence and
direction of each current component. Recall that, in our setup, the exchange
field in F1 is fixed along the $z$ axis (see Fig.~\ref{fig_model}) and therefore
contributes only to the generation of $J_x$. We first consider the $J_x$ profiles in
Fig.~\ref{switch_effect_weak_case}(c). At $\theta=90^\circ$, the exchange field
in F2 is directed entirely along the $x$ axis and therefore provides no
$z$-directed magnetic driving for $J_x$. The only such contribution then comes
from the much weaker exchange field in F1, resulting in an extremely small
$J_x$.

For $\theta\neq90^\circ$, a non-vanishing $J_x$ emerges, as expected.
The $J_x$ profiles at $\theta=0^\circ$ and $45^\circ$ have signs
opposite to those at $\theta=180^\circ$ and $135^\circ$, respectively. This
behavior reflects the reversal of the $z$ component of the exchange field in
F2. A similar angular dependence is observed for $J_z$. 
At $\theta=0^\circ$ and $180^\circ$, the $x$ component of the exchange 
field in F2 vanishes, and $J_z$ therefore vanishes identically.
At the remaining angles, the $x$ component of exchange field is positive, so the
resulting $J_z$ profiles have the same sign. These observations further
corroborate the conclusion that $h_2$ plays a substantially more important
role in the electromagnetic proximity effect in the weak-exchange-field
regime.


\begin{figure*}[ht!]
    \includegraphics[width=\textwidth,clip]{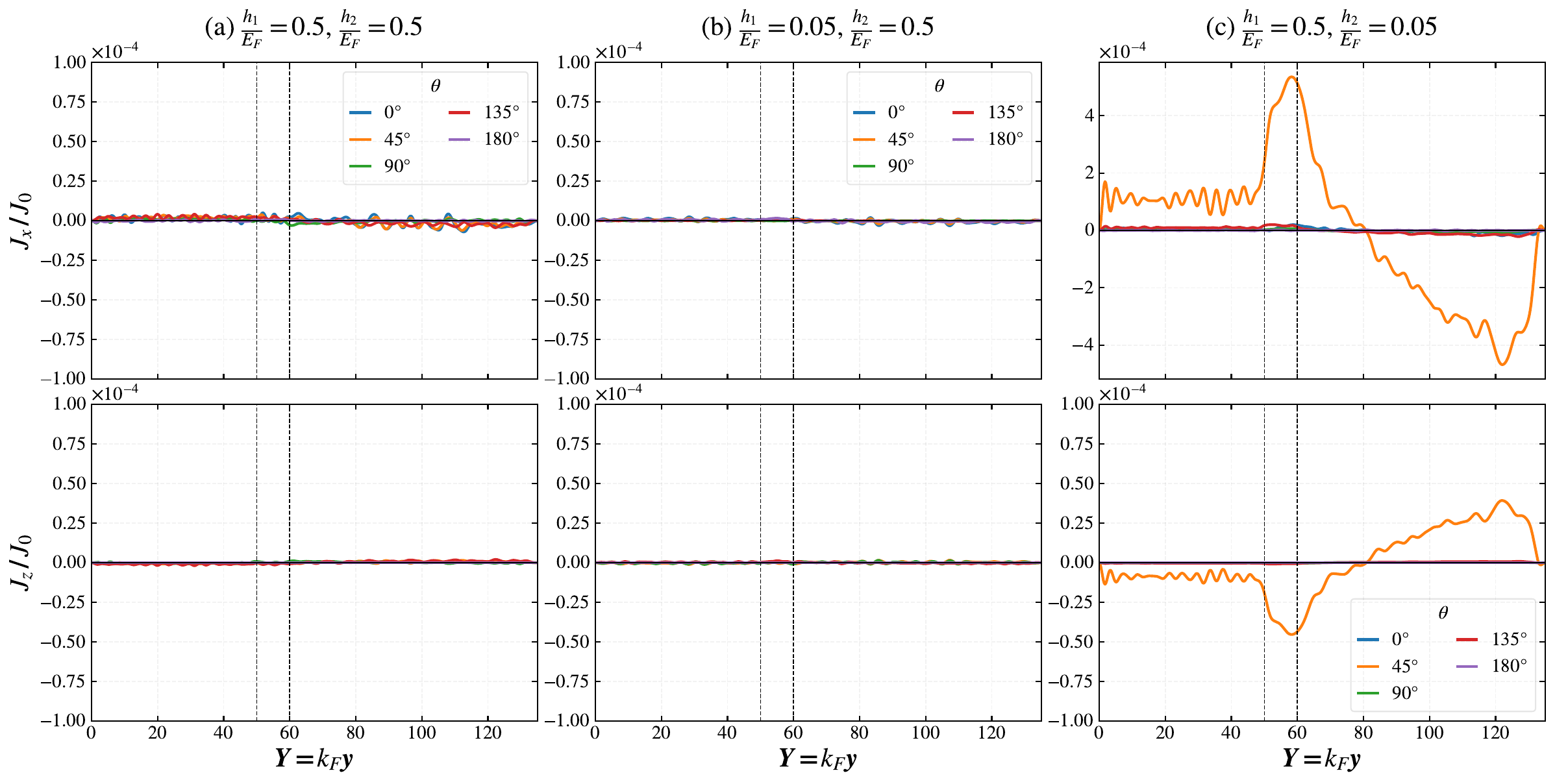}
    \caption{Spatial profiles of the spontaneous current density for three representative configurations in the strong-exchange-field regime. The results illustrate the switch-like role of the F2 layer in controlling the emergence of spontaneous currents. All parameters other than the exchange fields are fixed at the common values specified in the main text. The vertical dashed lines mark the interfaces between adjacent layers.}
    \label{switch_effect_strong_case}
\end{figure*}

\subsection{Regime of strong exchange field}

We next consider a stronger-exchange-field regime that also exhibits the
switch-like behavior dominated by the F2 layer. We first examine the case of
equal exchange-field strengths in F1 and F2, with
$h_1=h_2=0.5E_F$. Despite the substantially larger exchange fields, the
spontaneous current density remains negligible throughout the heterostructure for all
relative angles $\theta$, as shown in
Fig.~\ref{switch_effect_strong_case}(a). Although the stronger exchange fields
generate a local electromagnetic response near the ferromagnetic layers,
their influence does not penetrate sufficiently far into the superconducting
layer to sustain an appreciable current. Consequently, the current
redistribution required by Eq.~\eqref{current_flux_conservation} remains
insufficient to produce a non-negligible spontaneous current.

The exchange-field strength is next reduced in either the F1 or F2 layer, 
one at a time, with the corresponding results shown in Figs.~\ref{switch_effect_strong_case}(b) 
and \ref{switch_effect_strong_case}(c), respectively. It should be noted that 
the normalized current density fluctuations in Fig.~\ref{switch_effect_strong_case}(a) may appear 
larger than those in Fig.~\ref{switch_effect_strong_case}(b). 
However, their magnitudes are both of order $10^{-6}$, which, 
as discussed above, lies within the numerical uncertainty and 
therefore carries no physical significance. 
As shown in Fig.~\ref{switch_effect_strong_case}(b), 
reducing the exchange field in F1 from $h_1=0.5$ to $h_1=0.05$ produces 
almost no change in the spontaneous current density. This suggests 
that the strong exchange field in F2 strongly suppresses 
the superconducting proximity effect, thereby weakening 
the coupling between the superconducting and ferromagnetic layers 
and inhibiting the generation of spontaneous currents.

In contrast, Fig.~\ref{switch_effect_strong_case}(c) exhibits 
qualitatively different behavior when the exchange field in F2 
is reduced from $h_2=0.5$ to $h_2=0.05$. Although the spontaneous 
current remains negligible for most relative magnetization angles, 
an appreciable current distribution emerges at $\theta=45^\circ$. 
This result indicates that weakening the exchange field in F2 
partially restores the superconducting proximity effect and
enables the generation of spontaneous currents. It therefore 
further supports the switch-like role of the F2 layer in controlling 
the spontaneous-current response.

For the parameters considered in Fig.~\ref{switch_effect_strong_case}(c),
a pronounced spontaneous-current response is found at
$\theta=45^\circ$, whereas the current remains small at the other
sampled angles. The angle at which the current is enhanced is not
universal and varies with the microscopic parameters, reflecting the
competition among noncollinear triplet generation, proximity
penetration, and the angle-dependent contributions of current-carrying
states.

\begin{figure*}[ht!]
    \includegraphics[width=\textwidth,clip]{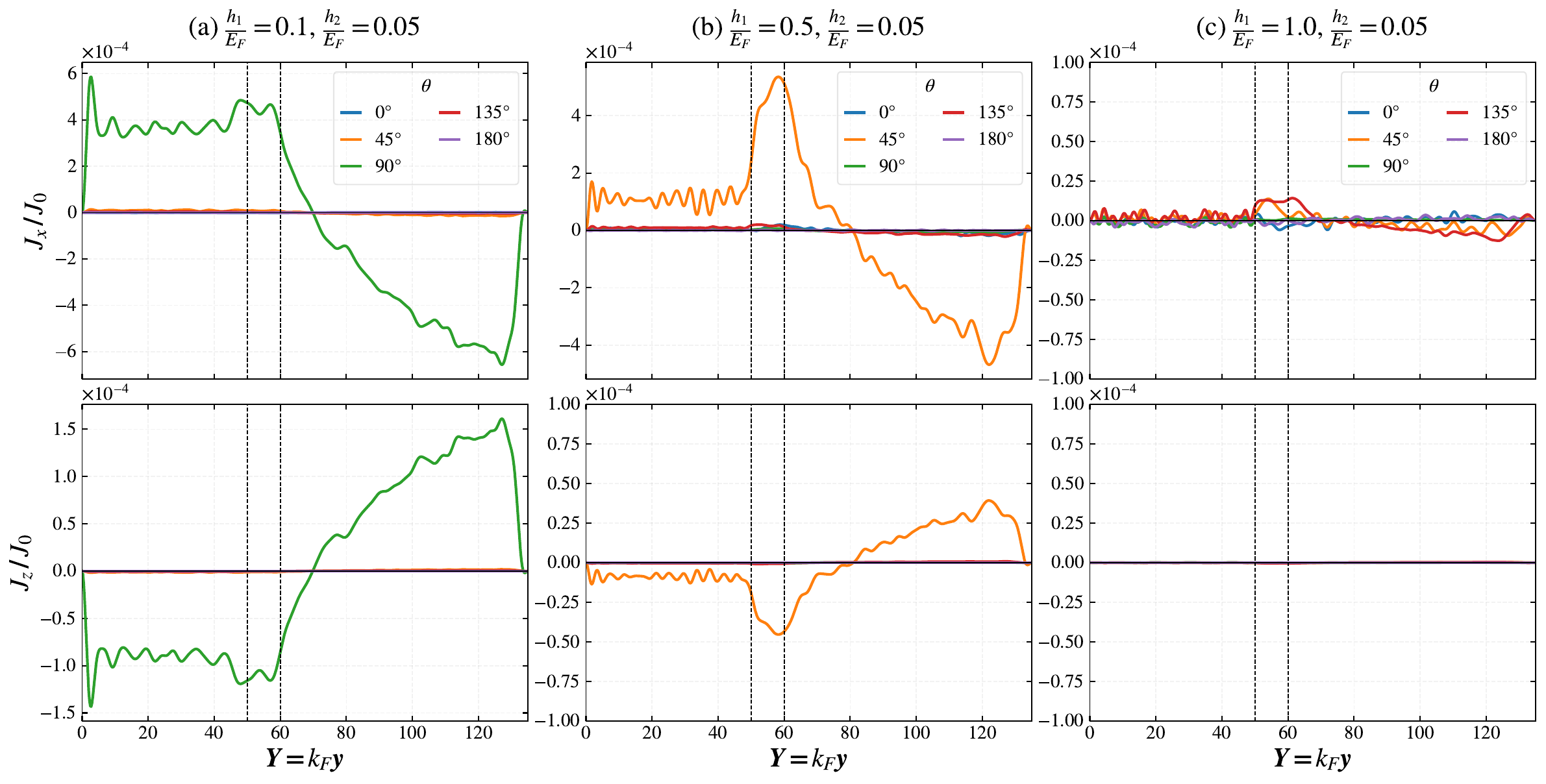}
    \caption{Spatial distributions of the spontaneous current in the heterostructure for fixed $h_2$ and different values of $h_1$, illustrating that $h_1$ modifies the current distribution even though the switch-like effect is primarily associated with the F2 layer.}
    \label{switch_effect_special}
\end{figure*}

\subsection{F1-controlled suppression regime}
In the above strong-exchange-field regime, the switch-like effect dominated by the F2 layer does not imply that the F1 layer is irrelevant in other parameter regimes. Signatures of the contribution from the F1 layer can be discerned in Fig.~\ref{switch_effect_special}. As the exchange field in the F1 layer is increased from $h_1=0.1$ to $h_1=0.5$ while $h_2$ is fixed at $0.05$, a clear change in the spatial distribution of the spontaneous current is observed. 


Here, rather than comparing the current-density profiles at a fixed relative magnetization angle, we focus on the profile with the largest amplitude for each value of $h_1$. The angle at which the spontaneous current is maximized is not universal and may occur at $\theta=45^\circ$, $90^\circ$, or $135^\circ$, depending on the microscopic parameters. 
The optimal angle for enhancing the current is again determined by the microscopic parameters, which govern the balance between the induced triplet correlations, the penetration of superconductivity, and the angular response of the current-carrying states.

Figure~\ref{switch_effect_special} clearly shows that increasing $h_1$ progressively modifies the spatial profile of the spontaneous current. In Fig.~\ref{switch_effect_special}(a), where $h_1=0.1$, the current extends throughout the F1 layer. When $h_1$ is increased to $0.5$, however, Fig.~\ref{switch_effect_special}(b) shows that the current penetrating into F1 is strongly reduced, while the largest current density appears in F2 and extends deep into the S layer. 
In the half-metallic limit, corresponding to a very large $h_1$, 
the spontaneous current is no longer discernible within F1 and 
survives only outside the F1 layer with a much smaller amplitude $J_x/J_0\approx 10^{-5}$, 
as shown in Fig.~\ref{switch_effect_special}(c).
The uncertainty associated with the initialization is far below
$10^{-6}$ in this case and is therefore negligible compared with the remaining spontaneous current.
Thus, a strong exchange field in F1 suppresses, in the half metallic limit, 
the overall magnitude of the spontaneous current throughout the heterostructure. This behavior can again be understood in terms of the superconducting proximity effect: as $h_1$ increases, the penetration of Cooper-pair correlations into F1 is strongly suppressed, thereby reducing the spontaneous current density.
Taking Figs.~\ref{switch_effect_strong_case} and~\ref{switch_effect_special}
together, we find that for $h_1\gtrsim0.1E_F$ an appreciable spontaneous
current appears only in noncollinear configurations, where long-ranged
triplet correlations are expected; for very weak $h_1$ [Fig.~\ref{switch_effect_weak_case}(c)], collinear
($\theta =0^\circ $, $\max\limits_{y\in S}\abs{J_x/J_0}\approx 8\times 10^{-4}$) and noncollinear ($\theta =90^\circ$, $\max\limits_{y\in S}\abs{J_z/J_0}\approx 7.5\times10^{-4}$)
configurations give currents of comparable magnitude.
We also find that, at $\theta=0^\circ$ and $h_1=h_2$, the electromagnetic proximity effect 
is usually negligible—or, equivalently, the spontaneous current 
is vanishingly small—over a wide range of 
exchange-interaction strengths, consistent with 
Ref.~\cite{Satchell_SupercondSciTechnol.36.054002}.

\subsection{Profiles of magnetic field}
\label{magnetic_field_profile}
\begin{figure*}[ht!]
    \includegraphics[width=\textwidth,clip]{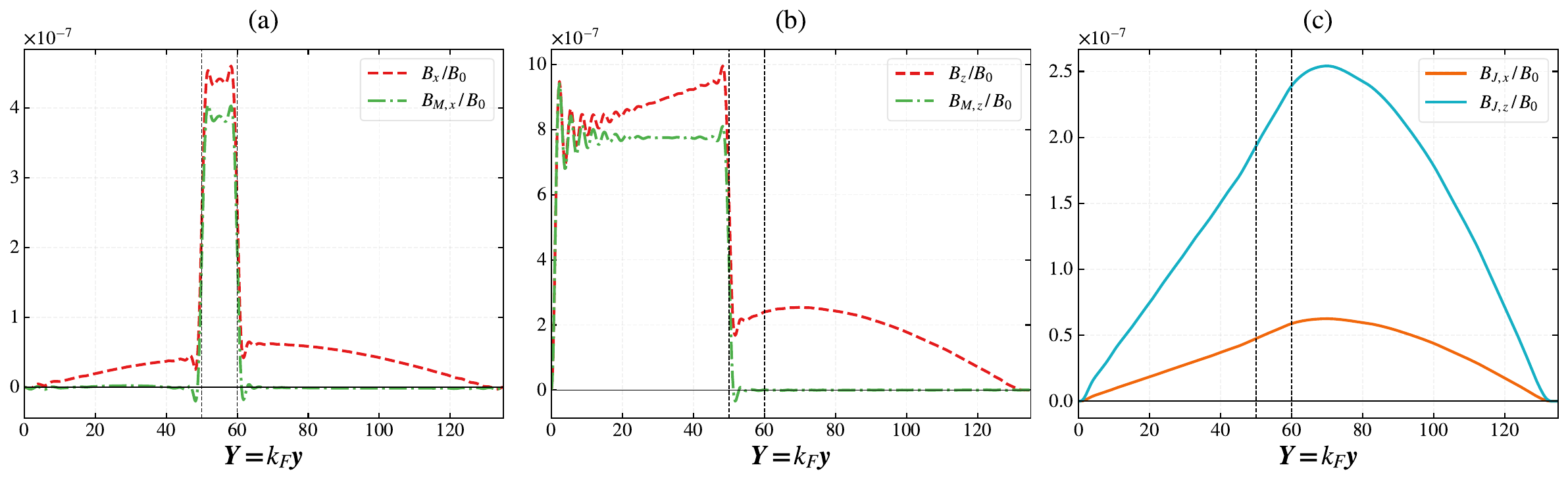}
    \caption{Magnetic-field profiles for the configuration corresponding to Fig.~\ref{switch_effect_special}(a) with $\theta=90^\circ$, showing the $x$- and $z$-components in panels (a) and (b), respectively. The total magnetic field ($\vb{B}$) is plotted together with the spin-induced contribution ($\vb{B}_{M}$) in panels (a) and (b). The $x$- and $z$-components of the current-induced magnetic field ($\vb{B}_{J}$) are shown in panel (c).}
    \label{magnetic field 1}
\end{figure*}
\begin{figure*}[ht!]
    \includegraphics[width=\textwidth,clip]{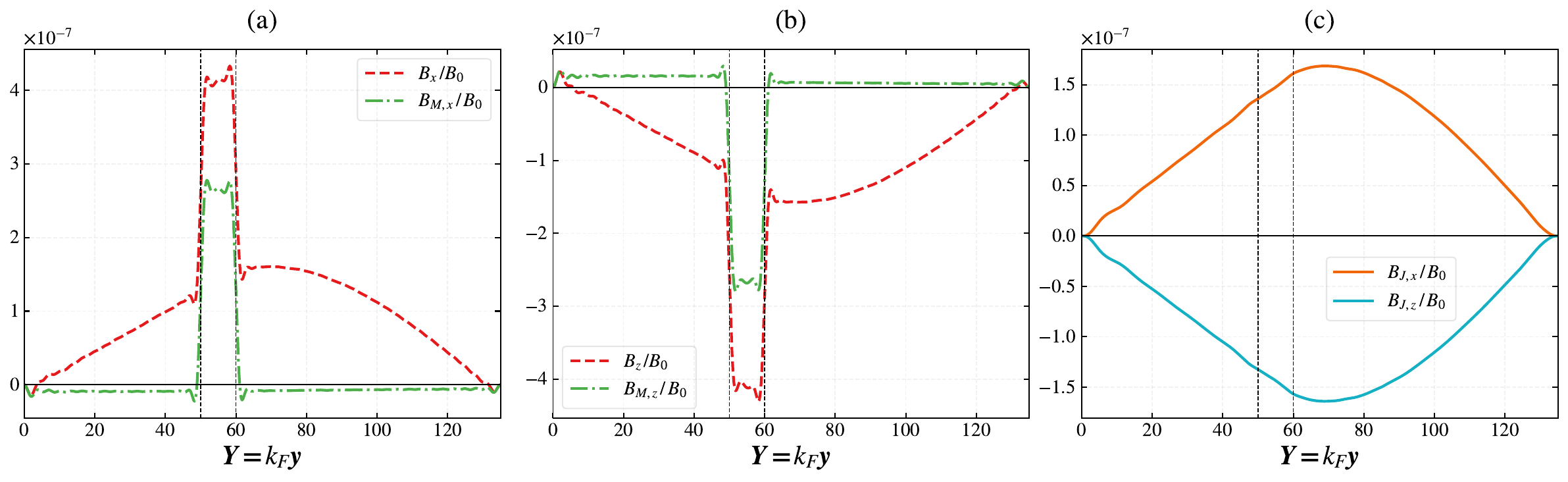}
    \caption{Magnetic-field distributions obtained for the parameter set used in Fig.~\ref{switch_effect_weak_case}(c) at $\theta=135^\circ$. Panels (a) and (b) display the $x$- and $z$-components, respectively, with the full magnetic field $\vb{B}$ compared against the spin-induced field $\vb{B}_{M}$. Panel (c) presents the corresponding $x$- and $z$-components of the field generated by the spontaneous current, $\vb{B}_{J}$.}
    \label{magnetic field 2}
\end{figure*} 

We now turn to the influence of spontaneous currents on the overall
magnetic-field distribution in the heterostructure. For convenience, we
normalize the magnetic field by  $B_0=\frac{\hbar c k_F^2}{|e|}$.
For the parameter values introduced at the beginning of
Sec.~\ref{Results}, this field scale is approximately
$B_0=4.9\times10^{8}\,\mathrm{G}$.

Figure~\ref{magnetic field 1} shows that, once the spontaneous current
becomes appreciable, its contribution to the magnetic field can be
comparable in magnitude to that generated by the spin magnetization.
We use the same parameter set as in
Fig.~\ref{switch_effect_special}(a), with the relative magnetization
angle fixed at $\theta=90^\circ$, for which a pronounced spontaneous
current is obtained. Figures~\ref{magnetic field 1}(a) and
\ref{magnetic field 1}(b) show the $x$- and $z$-components of the
magnetic field, respectively. In each panel, the total magnetic field,
which includes both the current- and spin-magnetization-induced
contributions, is compared with the field $\vb{B}_{M}$ generated solely
by the spin magnetization. The difference between the two curves
demonstrates that the correction arising from the spontaneous current
is appreciable.

The current-induced contribution $\vb{B}_{J}$ is shown separately in
Fig.~\ref{magnetic field 1}(c). Its magnitude is largest in the
vicinity of the F2/S interface and remains finite over a substantial
distance inside the superconducting layer. This spatial structure is
closely related to the current-conservation condition in
Eq.~\eqref{current_flux_conservation}, which requires compensating
current contributions flowing in opposite directions across the
heterostructure. Through Eqs.~\eqref{B fields}, these spatially
distributed currents generate the corresponding variation of
$\vb{B}_{J}$. In particular, the sign of each current component
determines the sign of the spatial gradient of the associated 
magnetic-field component. Importantly, the current-induced magnetic field
penetrates the entire superconducting layer,
consistent with the behavior reported experimentally in
Refs.~\cite{Flokstra_ApplPhysLett.115.072602,
Stewart_PhysRevB.100.020505}.


An additional instructive case is obtained using the parameters of
Fig.~\ref{switch_effect_weak_case}(c) with the relative magnetization
angle fixed at $\theta=135^\circ$. The corresponding magnetic-field
profiles are shown in Fig.~\ref{magnetic field 2}. Because
$h_1=0.001\,E_F$, the spin-magnetization-induced field $\vb{B}_M$ is
negligibly small throughout most of the heterostructure, except within
the F2 layer. Consistent with the switch-like behavior discussed above,
F2 therefore plays the dominant role in generating the spontaneous
current in this weak-exchange-field regime, particularly when $h_1$ is
very small. Consequently, the current-induced field $\vb{B}_J$ substantially alters
the total magnetic field in the F1 layer. In particular, it generates
field components whose signs and relative magnitudes are governed by
the magnetization orientation of F2, thereby reorienting the net field
away from that expected from the weak F1 magnetization alone, as shown
in Figs.~\ref{magnetic field 2}(a) and \ref{magnetic field 2}(b).
Interestingly, $\vb{B}_J$ can become comparable in magnitude to $\vb{B}_M$, 
indicating that the electromagnetic proximity effect provides an effective 
mechanism for reconfiguring the local magnetic-field distribution
in the heterostructure~\cite{Kopasov_PhysRevB.110.214501,Kopasov_JSupercondNovMagn.38.241}.
Finally, Fig.~\ref{magnetic field 2}(c) also demonstrates that the electromagnetic
proximity effect is relatively long-ranged, at least in terms of its
influence on the magnetic-field profile. Once an appreciable
spontaneous current develops, the associated field $\vb{B}_J$ extends
throughout the heterostructure.

\begin{figure*}[ht!]
    \includegraphics[width=\textwidth,clip]{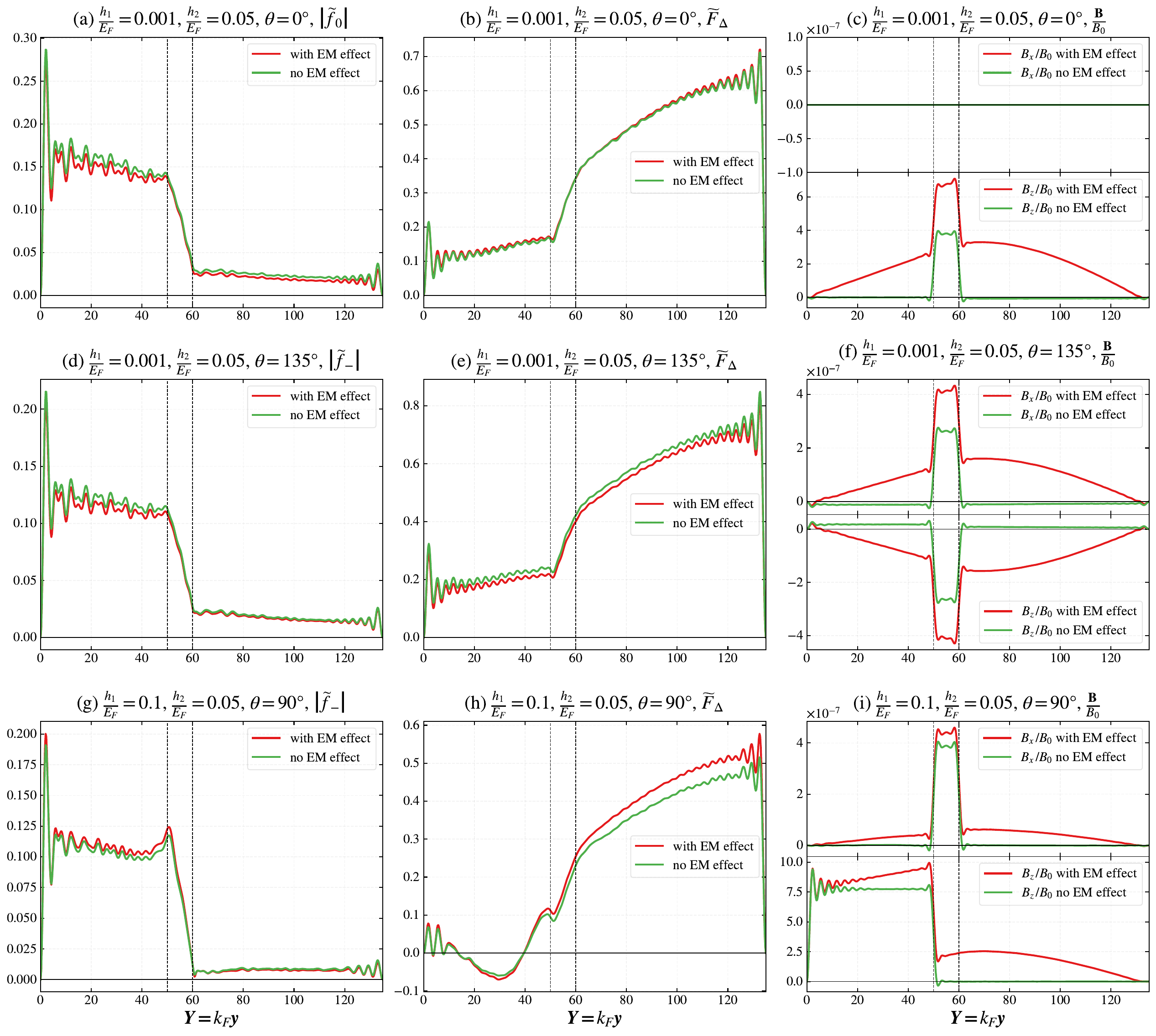}
    \caption{Representative cases for the parameter sets of Figs.~\ref{switch_effect_weak_case}(c) and \ref{switch_effect_special}(a). Spatial distributions of the normalized triplet component $\big|\widetilde{f}_0\big|$ or $\big|\widetilde{f}_{-}\big|$, the normalized singlet amplitude $\widetilde{F}_{\Delta}$, and the magnetic-field components $\vb{B}/B_0$ are shown from left to right. The upper [(a)--(c)] and middle [(d)--(f)] panels correspond to the same parameters as Fig.~\ref{switch_effect_weak_case}(c), with $\theta=0^\circ$ and $135^\circ$, respectively. The lower panels [(g)--(i)] correspond to the same parameters as Fig.~\ref{switch_effect_special}(a), with $\theta=90^\circ$. Panel (a) shows $\big|\widetilde{f}_0\big|$, whereas panels (d) and (g) show $\big|\widetilde{f}_{-}\big|$. In the right column, the upper and lower subpanels show $B_x/B_0$ and $B_z/B_0$, respectively. Red (green) curves represent results with (without) the electromagnetic proximity effect. Note that $\theta=0$, $B_x/B_0=0$ in (c). The red curves in panel (f) and (i) are reproduced from Figs.~\ref{magnetic field 2}[(a)-(b)] and Figs.~\ref{magnetic field 1}[(a)-(b)] for comparison, respectively.}
    \label{triplet_all_new}
\end{figure*}

\subsection{Profiles of odd-frequency spin-triplet correlations}
Finally, we turn to the influence of the spontaneous current on odd-frequency spin-triplet correlations, which constitute one of the most important features of superconducting heterostructures.
Although these correlations do not themselves correspond directly to a physical observable, our numerical results show that their behavior is correlated with the emergence of spontaneous currents. In the following discussion, the correlation functions $f_{0}$ and $f_{\pm}$ are normalized by ${\Delta_{\text{bulk}}}/{\mathcal{V}_{\text{eff}}}$ and denoted by $\widetilde{f}_{0}$ and $\widetilde{f}_{\pm}$, respectively. Here, $\mathcal{V}_{\text{eff}}$ is the BCS singlet coupling constant, which can be estimated as $\mathcal{V}_{\text{eff}}={\Lambda}/{d_{\sigma}(E_F)}$ where $\Lambda=\left[\ln(R+\sqrt{R^2+1})\right]^{-1}, R={\hbar\omega_D}/{\Delta_{\text{bulk}}}$, and $d_{\sigma}(E_F)=\frac{m^* k_F}{2\pi^2\hbar^2}$ is the density of states at the Fermi energy per unit volume and per spin. In addition, the time dependence in Eqs.~\eqref{triplet correlations_a} and \eqref{triplet correlations_b} is exhibited in oscillatory factors $\sin(\frac{E_n}{\hbar}t)$ and $\cos(\frac{E_n}{\hbar}t)$. For notational convenience, the dimensionless time $\tau=\omega_D t$ is introduced, so that their common phase becomes $\frac{E_n}{\hbar\omega_D}\tau$. Since the following analysis is not sensitive to the specific choice of $\tau$, $\tau=4$ (corresponding to $t=4\omega_D^{-1}$) is used throughout the rest of the paper to avoid the overly early-time regime. Finally, the singlet pairing amplitude $F_{\Delta}(y)$ is also expressed in units of $\Delta_{\text{bulk}}/\mathcal{V}_{\text{eff}}$ to facilitate visualization of its spatial profile, with the resulting dimensionless quantity denoted by $\widetilde{F}_{\Delta}$.\par

In the following discussion, we examine the influence of spontaneous-current configurations on the odd-frequency spin-triplet correlations by comparing simulations with and without electromagnetic proximity effects.
For the latter, we impose $\vb{A}=0$ at every iteration step, 
rather than updating it self-consistently from the Maxwell equations, 
and simultaneously omit the magnetic-field-dependent Zeeman interaction.
When $\vb{A}=0$, it is straightforward to see that the states 
are symmetric under $\vb{k}_\perp\mapsto-\vb{k}_\perp$, implying that 
the current density vanishes, as is also evident from Eq.~\eqref{J_i}.
In other words, after fixing the physical parameters of the system, 
the iterative procedure is restricted to the self-consistent determination of $\Delta(y)$, 
without any feedback from Maxwell's equations.
Although we enforce $\vb{A}=0$, thereby turning off the influence of magnetic field
in this procedure, this does not mean 
that the magnetic field is zero. Instead, it is directly proportional to
the self-consistent magnetization, i.e., $\vb{B}=\vb{B}_M=4\pi \vb{M}$. 
We note that these two procedures should not be expected to 
yield the same \(\vb{B}_M\), since the former retains the 
influence of \(\vb{A}\) at each iteration step, whereas the latter does not.
The examples discussed below allow us to summarize 
several simple trends and provide a qualitative 
phenomenological picture of the underlying behavior.\par

First, we investigate the cases in which $h_1$ and $h_2$ are collinear. Here, the only nonvanishing triplet correlation is $\widetilde{f}_0$, since nonzero $\widetilde{f}_{\pm}$ is not allowed from the symmetry consideration as discussed in Sec.~\ref{intro}.
We adopt the parameter set of Fig.~\ref{switch_effect_weak_case}(c) and consider the case $\theta=0^\circ$. Figure~\ref{triplet_all_new}(a) shows that including the electromagnetic proximity effect, and hence the spontaneous current, leads to a slight reduction in $\big|\widetilde{f}_0\big|$. In contrast, Fig.~\ref{triplet_all_new}(b) shows no noticeable modification of $\widetilde{F}_{\Delta}$. 

To relate the slight reduction in $\big|\widetilde{f}_0\big|$ to the spontaneous-current distribution, we plot the $z$ component of the magnetic field at $\theta=0^\circ$ in Fig.~\ref{triplet_all_new}(c). Recall that, in the absence of the electromagnetic proximity effect, the magnetic field originates solely from the magnetization. Figure~\ref{triplet_all_new}(c) therefore shows that the field $\vb{B}_J$ generated by the spontaneous current extends over much of the F1 and F2 regions and enhances the local magnetic field as in the discussion in Sec.~\ref{magnetic_field_profile}. 
We emphasize that the Zeeman effect from the magnetic field is
negligible compared to that from the exchange fields in the F layers.
Therefore, the influence of the resultant magnetic field 
enters through the vector potential $\vb{A}$ in the kinetic energy term in Eq.~\eqref{Hamiltonian}.
Through the associated orbital coupling, this additional electromagnetic response 
modifies the quasiparticle states and hence the superconducting correlations 
in the ferromagnetic layers. The resulting change is manifested as 
the slight reduction in $\big|\widetilde{f}_0\big|$ 
when the electromagnetic proximity effect is included.\par


Next, we consider noncollinear configurations of $h_1$ and $h_2$. 
Owing to the reduced symmetry in the noncollinear configurations, the $\widetilde{f}_{\pm}$ triplet components are allowed. Since these components are characteristic quantities for noncollinear cases, 
we focus on $\widetilde{f}_{\pm}$ in the following discussion, even though $\widetilde{f}_0$ remains finite.
As in the collinear cases, we retain the parameter set of Fig.~\ref{switch_effect_weak_case}(c) and examine the noncollinear configuration with $\theta=135^\circ$.
The relevant quantities are plotted in Figs.~\ref{triplet_all_new}[(d)-(f)].
As shown in Figs.~\ref{triplet_all_new}(d) and \ref{triplet_all_new}(e), 
the inclusion of the electromagnetic proximity effect, 
accompanied by the emergence of spontaneous currents, 
leads to a slight overall reduction in both $\big|\widetilde{f}_{-}\big|$ and $\widetilde{F}_{\Delta}$.
Figure~\ref{triplet_all_new}(f) shows the corresponding magnetic-field profile. 
Consistent with the previously discussed long-ranged electromagnetic proximity effect,
$\vb{B}_{J}$ extends throughout the heterostructure and, as noted in Sec.~\ref{magnetic_field_profile}, 
shifts the total magnetic-field orientation in F1 toward that in F2, 
owing to the very weak exchange field $h1=0.001E_F$.

The analysis is next extended to another noncollinear configuration, 
using the parameter set of Fig.~\ref{switch_effect_special}(a) at $\theta=90^{\circ}$.
The relevant quantities are presented in Figs.~\ref{triplet_all_new}[(g)-(i)]. 
In contrast to the preceding case, the inclusion of the electromagnetic proximity effect 
associated with spontaneous-current generation leads to an overall enhancement of 
both $\big|\widetilde{f}_{-}\big|$ and $\widetilde{F}_{\Delta}$, as shown in Figs.~\ref{triplet_all_new}(g) and \ref{triplet_all_new}(h).
Figure~\ref{triplet_all_new}(i) again shows that $\vb{B}_J$ extends throughout the heterostructure. 
Because $h_1$ and $h_2$ are comparable and neither is very weak, 
the total field in each F layer deviates from its respective exchange-field direction, rather than being dominated by the other layer.
While the direction of $\vb{B}$ varies more strongly in the noncollinear cases than in the collinear case, 
its Zeeman contribution is still much smaller than that of the exchange field. 
Accordingly, the modifications of $\widetilde{F}_{\Delta}$ and $\big|\widetilde{f}_{-}\big|$ induced by the electromagnetic proximity effect 
should still arise predominantly from the orbital coupling through $\vb{A}$.

Next, we provide a conceptual explanation of why 
the orbital coupling through \(\vb{A}\) can modify 
the correlation functions of the system, even though 
the Zeeman effect associated with the electromagnetic 
proximity effect is negligible compared with 
the exchange-field contribution.
First, the exchange interaction and the Zeeman coupling to $\vb{B}$ essentially 
split the spin-dependent energy bands, with the relevant spin-quantization axis.
In momentum space, this produces different $\vb{k}_{\perp}$ contours for the two spin branches at a given energy. 
This means that, at a given energy, the
two spin branches have circular contours with different radii in $\vb{k}_{\perp}$ space. 
Because the Zeeman effect of $\vb{B}$ is negligible relative to the exchange interaction, 
the difference in circular-contour radii between the two spin branches is controlled almost entirely by the exchange field.
However, the electromagnetic proximity effect can still influence the system through the vector potential $\vb{A}$ entering the orbital coupling.
This behavior is more transparent in the momentum space of Nambu spinor. 
The particle-like and hole-like sectors couple to $\vb{A}$ with opposite signs through the orbital term, 
leading schematically to the shifts $\vb{k}_{\perp}\mapsto\vb{k}_{\perp}+\frac{\abs{e}}{\hbar c}\vb{A}$ and $\vb{k}_{\perp}\mapsto\vb{k}_{\perp}-\frac{\abs{e}}{\hbar c}\vb{A}$, respectively.
Equivalently, the presence of $\mathbf{A}$ shifts the centers of the circular contours in the two sectors 
by equal magnitudes in opposite directions relative to the $\mathbf{A}=0$ case, i.e., the case without the electromagnetic proximity effect.
Clearly, the Zeeman effect and the orbital coupling influence the formation of the correlation functions through fundamentally different mechanisms: 
the former modifies the radii of the circular contours of different spin branches, 
whereas the latter shifts the centers of the circular contours in the particle- and hole-like sectors.
This explains why the electromagnetic proximity effect can affect the correlation functions even though its Zeeman contribution is almost negligible. 
In contrast to the case without the electromagnetic proximity effect, 
its inclusion introduces a qualitatively distinct mechanism that reshapes the circular contours
thereby modifying $\widetilde{F}_{\Delta}, \widetilde{f}_0$ and $\widetilde{f}_{-}$.

With the above mechanism in mind, a qualitative interpretation of the collinear and noncollinear cases can be provided. 
In the collinear case, even after the electromagnetic proximity effect is included, 
the induced magnetic field $\vb{B}$ remains collinear with the exchange field along the $z$ direction. 
It is therefore reasonable to expect that the orbital coupling through $\mathbf{A}$ produces collinear shifts of the 
constant energy circular contours, 
with only the magnitude and sign of the shift varying with $y$. 
Consequently, the corrections to $\langle\psi_{\uparrow}(\mathbf{r},t)\psi_{\downarrow}(\mathbf{r},0)\rangle$ and $\langle\psi_{\downarrow}(\mathbf{r},t)\psi_{\uparrow}(\mathbf{r},0)\rangle$ may approximately have equal magnitudes but opposite signs. 
Such a cancellation could account for the nearly unchanged $\widetilde{F}_{\Delta}$ and the slight overall reduction of $\big|\widetilde{f}_0\big|$ in Figs.~\ref{triplet_all_new}[(a)-(b)].

The situation is considerably more complicated in the noncollinear cases. 
Here, not only does the exchange-field orientation change between the F layers, 
but the magnetic field generated by the electromagnetic proximity effect also varies in both magnitude and direction with $y$ 
and need not remain aligned with the local exchange field. 
The resulting orbital coupling can therefore produce $y$-dependent shifts 
of the centers of the circular momentum-space contours, 
with both the magnitude and direction of the shifts varying with $y$.
In addition, because the exchange-field directions differ between F1 and F2, a $z$-axis spin state is generally a linear combination of the two spin branches defined by the local exchange field in the F2 layer. 
Consequently, both corresponding contours must be considered simultaneously.
These considerations may explain why the electromagnetic proximity effect produces complex behavior in Figs.~\ref{triplet_all_new}[(d)–(e)] and Figs.~\ref{triplet_all_new}[(g)–(h)]. 
In the noncollinear case, the orbital and spin-splitting mechanisms coexist and jointly modify, 
in a highly complex manner, the phase space available for the formation
of $\widetilde{F}_{\Delta}$ and $\widetilde{f}_-$. 
Finally, we stress that the mechanism discussed above 
should be regarded only as a qualitative interpretation 
and not as a universal, quantitatively exact description.

\section{Conclusion}
In this work, we investigate the influence of electromagnetic fields on various proximity effects in clean ferromagnet/ferromagnet/superconductor spin-valve heterostructures. We develop a numerical scheme that solves the Bogoliubov–de Gennes and Maxwell equations simultaneously and self-consistently. The two sets of equations are coupled through a self-consistently determined vector potential, from which the current density and magnetic-field profile are obtained. In contrast to previous studies of ferromagnet/superconductor heterostructures based on quasiclassical methods, the BdG approach provides a microscopic description and is well suited for resolving physics on atomic length scales.
Importantly, our formalism does not invoke the phenomenological London relation as an additional assumption and is therefore fully self-contained. 

We predict that spontaneous currents can arise in spin-valve heterostructures 
when electromagnetic proximity effects are properly taken into account. 
In the weak-exchange-field regime, the thin central F2 layer plays a more prominent role 
than F1 in controlling the current response. Starting from a configuration 
in which both exchange fields are very weak and the current density is essentially zero, 
increasing the exchange field in F2 generates a sizable spontaneous current, 
whereas a comparable increase in F1 does not. A similar asymmetry is found 
in the stronger-exchange-field regime: reducing the exchange field in F2 
from a symmetric strong-field configuration produces a pronounced spontaneous current, 
while the corresponding reduction in F1 has little effect. These results highlight 
the distinct roles of the two ferromagnetic layers in determining the spontaneous-current response.

Guided by these observations, we fix \(h_2=0.05E_F\), for which the spontaneous current 
is appreciable, and vary \(h_1\). As \(h_1\) is reduced from \(0.5E_F\) to \(0.1E_F\), 
the relative angle at which the current density is maximized
(among the sampled angles $\theta=0^\circ,\,45^\circ,\,90^\circ,\,135^\circ,\,180^\circ$)
shifts from \(45^\circ\) to \(90^\circ\), 
while the magnitude of maximum current density is not decreased appreciably. 
By contrast, upon approaching the half-metallic limit, 
the spontaneous current is strongly suppressed, presumably because of the weakened superconducting proximity effect.
We also investigate the magnetic-field profiles in cases 
where the spontaneous current is appreciable. 
We find that the current-induced contribution 
to the magnetic field is comparable in magnitude to that 
arising from the spin magnetization, indicating that 
electromagnetic proximity effects cannot be neglected 
when analyzing the magnetic response of the heterostructure. 
The current-induced magnetic field extends across
the superconducting layer, with its maximum occurring 
near the F2/S interface, demonstrating the long-ranged nature 
of the electromagnetic proximity effect. We further find that 
the electromagnetic proximity effect tends to reconfigure  
the local magnetic-field orientation in the spin-valve heterostructure 
so as to reduce the angular mismatch between the magnetic fields 
in the two ferromagnetic layers.

Next, we examine the influence of electromagnetic proximity effects 
on the singlet and odd-frequency triplet amplitudes. We find that, 
when electromagnetic proximity effects are included, 
both quantities are only weakly modified. 
This suggests that if one is primarily interested 
in the pairing amplitudes themselves, 
rather than in the magnetic response, electromagnetic effects 
may not play a crucial role. In contrast, an accurate description 
of the magnetic-field profile requires the electromagnetic proximity effect 
to be taken into account, since it can substantially reconfigure 
the local magnetic-field orientation and reduce the angular mismatch 
between the fields in the two ferromagnetic layers. 
Overall, the electromagnetic proximity effect modifies the superconducting 
correlations mainly through the orbital coupling to $\vb{A}$, which shifts
the spin-dependent constant-energy contours in momentum space. 
These changes feed back into both the singlet 
and triplet pair amplitudes, although the resulting corrections are generally not large.

Our numerical scheme provides a general framework 
for treating electromagnetic effects in superconducting heterostructures. 
In particular, it can possibly be extended to study current-phase relations 
in superconductor/ferromagnet/superconductor Josephson junctions. 
The same framework can also be applied to investigate Meissner screening 
and other self-consistent electromagnetic responses in such systems.
Possible future directions include applications 
to superconducting diode effects~\cite{PhysRevB.105.104508} and to heterostructures 
incorporating altermagnets~\cite{altermanget}, where the interplay 
between unconventional magnetism and electromagnetic proximity effects 
may lead to additional functionalities.

\begin{acknowledgments}
This work was supported by the National Science and Technology Council (NSTC) of Taiwan under Grant No.~NSTC 114-2112-M-A49-019-. C.-T.~W. acknowledges additional support from the Center for Theoretical and Computational Physics (CTCP), National Yang Ming Chiao Tung University. 
\end{acknowledgments}

\section*{Data availability}
The data that support the findings of this article are not publicly available. 
The data, and the code used to generate them, 
are available from the authors upon reasonable request.

\bibliography{ref}

\end{document}